\documentclass[prd,twocolumn,showpacs,floatfix,amsmath,nofootinbib,amssymb,floatfix]{revtex4}
\usepackage{graphicx,color,dcolumn,booktabs,bm}
\usepackage{longtable,lscape}
\usepackage{txfonts}
\usepackage{overpic}
\usepackage{amssymb}
\usepackage{indentfirst}
\usepackage{feynmf}   
\usepackage{slashed}  
\usepackage{cases}
\usepackage{color}
\usepackage{multirow}
\usepackage{epstopdf}
\usepackage{longtable}
\usepackage{float}
\usepackage{graphicx,color,dcolumn,booktabs,bm}
\usepackage[colorlinks,
            citecolor=blue,
            anchorcolor=red,
            menucolor=red,
            linkcolor=red,
            filecolor=red,
            runcolor=red,
            urlcolor=blue,
            frenchlinks=red]{hyperref}

\graphicspath{{Figures}} %

\allowdisplaybreaks

\begin{document}

\title{Molecular pentaquarks composed of a ground-state octet baryon and a $P-$wave anticharmed meson}

\author{Yu-Yue Cui$^{1}$}
\author{Rui Chen$^{1,2}$}\email{chenrui@hunnu.edu.cn}
\author{Qi Huang$^{3}$}

\affiliation{
$^1$Department 
of Physics and 
Synergetic 
Innovation Center 
for Quantum Eﬀects
and Applications,
Key Laboratory of Low-Dimensional Quantum Structures and Quantum Control of Ministry of Education,  Hunan Normal University, Changsha 410081, China\\
$^2$Hunan Research Center of the Basic Discipline for Quantum Effects and Quantum Technologies, Hunan Normal University, Changsha 410081, China\\
$^3$School of Physics and Technology, Nanjing Normal University, Nanjing 210023, China}
\date{\today}

\begin{abstract}

In this work, we investigate the interactions between an excited
anticharm  meson doublet $(\bar{D}_1, \bar{D}_2^*)$ and ground-state octet baryons $(N, \Lambda, \Sigma, \Xi)$ with the aim of identifying possible molecular pentaquark states. A systematic analysis is performed within the one-boson-exchange model, which incorporates both $S$ and $P$-wave interactions, $S$-$D$ mixing, and coupled-channel effects. By solving the Schr\"{o}dinger equations, we can predict a rich spectrum of loosely bound anticharm  molecular pentaquarks with strangeness $|S| = 0, 1, 2$. Our results provide specific quantum number assignments and mass range predictions to guide future experimental searches at facilities such as LHCb and Belle II. The discovery of such states would significantly enrich the hadron spectrum and serve as a critical test of theoretical models for hadronic interactions.

\end{abstract}


\maketitle

\section{introduction}

The search for exotic states—hadrons beyond the conventional quark-antiquark mesons and three-quark baryons—has been a central pursuit in hadron physics since the quark model was proposed \cite{Gell-Mann:1964ewy,Zweig:1964ruk,Zweig:1964jf}. These exotic structures serve as unique laboratories for probing quantum chromodynamics in the strongly coupled, nonperturbative regime, advancing our fundamental understanding of matter and interaction.

Among these particles, pentaquark states, comprising four quarks and one antiquark, constitute an important and intriguing category of exotic hadrons. Theoretical investigations on their existence can be traced back to 1979, when Strottman calculated the masses of light-quark pentaquarks within the MIT bag model \cite{Strottman:1979qu}. Then, studies that explored the possibilities of stable pentaquarks containing a charm quark were carried out \cite{Lipkin:1987sk,Gignoux:1987cn}. The first tentative experimental evidence emerged in 2003 with the LEPS Collaboration's report of the $\Theta(1540)$, a state interpreted as a pentaquark composed of four light quarks and an $\bar{s}$ antiquark \cite{LEPS:2003wug}. However, subsequent high-statistics experiments failed to confirm this observation \cite{Hicks:2012zz}. Similarly, early evidence for other candidates, such as the double-strange $\Xi^{--}(ddss\bar{u})$ \cite{NA49:2003fxh} and the anticharmed $\Theta_c(uudd\bar{c})$ \cite{H1:2004gdk}, could not be independently verified \cite{Fischer:2004qb,ZEUS:2004lsr,WA89:2004bmh}.

However, things changed dramatically in the past decade. In 2015, the LHCb Collaboration announced the discovery of two hidden-charm pentaquarks, $P_c(4380)$ and $P_c(4450)$, in the decay $\Lambda_b^0 \to J/\psi p K^-$ \cite{LHCb:2015yax}. A subsequent analysis with combined Run I and Run II data in 2019 not only revealed a new narrow state, $P_c(4312)$, but also resolved the $P_c(4450)$ into two distinct peaks, $P_c(4440)$ and $P_c(4457)$ \cite{LHCb:2019kea}.

The proximity of these states to the $\Sigma_c \bar{D}^{(*)}$ thresholds has led to a widespread molecular interpretation in theoretical studies \cite{Du:2021fmf,Shen:2024nck,Wang:2019spc,Zhu:2019iwm,Burns:2019iih,Gutsche:2019mkg,Yamaguchi:2019seo,Wang:2019hyc,Voloshin:2019aut,Meng:2019ilv,Wang:2019ato,Guo:2019kdc,Chen:2019bip,Du:2019pij,Fernandez-Ramirez:2019koa,Yan:2022wuz,Wang:2022mxy,Dong:2021juy,Xiao:2019aya,Xiao:2019mvs,Chen:2016heh,Chen:2021cfl,Ke:2023nra,Karliner:2022erb,Lin:2019qiv,Chen:2019asm,Liu:2019tjn,He:2019ify,Xu:2020gjl}. In our previous work, we employed the one-boson-exchange (OBE) model to study the interactions between a charmed baryon and an anticharmed meson. Our results supported the identification of $P_c(4312)$, $P_c(4440)$, and $P_c(4457)$ as $\Sigma_c\bar{D}$, $\Sigma_c\bar{D}^*$, and $\Sigma_c\bar{D}^*$ molecules with $I(J^P)=1/2(1/2^-)$, $1/2(1/2^-)$, and $1/2(3/2^-)$, respectively, and highlighted the importance of coupled-channel effects \cite{Chen:2019asm}. It is worth noting that theoretical predictions for such hidden-charm molecular baryons existed prior to their experimental discovery \cite{Molina:2012mv,Yang:2011wz,Wang:2011rga,Karliner:2015ina,Wu:2010jy,Li:2014gra}.

Obviously, with the ongoing accumulation of experimental data and advances in technique, discoveries on additional pentaquark states must be possible. Thus, it is both timely and important to carry out theoretical investigations on other possible molecular pentaquark configurations. Such studies can not only enrich the hadron spectrum, but also provide a deeper understanding of strong interactions in the nonperturbative region, and offer further insights into the molecular nature of the observed exotic states.

In this work, we investigate the interactions between an anticharmed meson $\bar{T}=(\bar{D}_1, \bar{D}_2^*)$ and an octet baryon $B=(N, \Lambda, \Sigma, \Xi)$ to predict possible $\bar{T}B$ molecular pentaquarks. Compared to the well-studied $\Sigma_c\bar{D}^{(*)}$ systems, the $\bar{T}B$ configurations present distinct characteristics. On the one hand, the larger number of light quarks in the $\bar{T}B$ systems may enhance the contributions from light meson exchange. On the other hand, the relatively smaller reduced mass of these systems results in a larger kinetic energy, which can suppress bound-state formation. Thus, these two opposite traits make such a kind of system interesting and meaningful to be investigated, which can at least help us seek the boundary of interaction through experimental verification.

Our exploration builds upon earlier work, where we predicted possible molecular pentaquarks composed of $D_1N$ and $D_2^*N$ using an effective potential based on one-pion exchange \cite{Chen:2014mwa}. The $\bar{T}B$ systems offer a simpler theoretical framework than their $T B$ counterparts, as they avoid complications arising from quark annihilation processes. In the present study, we employ a comprehensive OBE model to derive the effective potentials, which means apart from the long-range OPE contribution, we also include intermediate- and short-range interactions mediated by scalar ($\sigma$) and vector ($\rho$, $\omega$) meson exchanges. In addition, we consider the coupled-channel effects, which were previously shown to be crucial for generating $D_1N$ and $D_2^*N$ molecular candidates \cite{Chen:2014mwa}. The coupled-channel Schrödinger equation is then solved to search for loosely bound molecular states.

This paper is organized as follows. In Sec.~\ref{sec2}, we derive the OBE effective potentials for the systems under study. The numerical results are presented and discussed in Sec.~\ref{sec3}. Finally, a summary is provided in Sec.~\ref{sec4}.

\section{The OBE interactions}\label{sec2}

We begin with the constructions on the flavor and spin-orbit wave functions, which are summarized in Table \ref{flavor wave}. The general forms of the spin-orbit wave functions are
\begin{eqnarray}
|\bar{D}_1B({}^{2S+1}L_{J})\rangle &=& \sum_{m,m',m_Sm_L}C^{S,m_S}_{\frac{1}{2}m,1m'}C^{J,M}_{Sm_S,Lm_L}
          \chi_{\frac{1}{2}m}\epsilon^{m'}|Y_{L,m_L}\rangle,\label{001}\nonumber\\
|\bar{D}_2^*B({}^{2S+1}L_{J})\rangle &=& \sum_{m,m'',m_Sm_L}C^{S,m_S}_{\frac{1}{2}m,2m''}C^{J,M}_{Sm_S,Lm_L}
          \chi_{\frac{1}{2}m}\zeta^{m''}|Y_{L,m_L}\rangle,\label{002}\nonumber
\end{eqnarray}
where $C^{J,M}_{j_1 m_1, j_2 m_2}$ are Clebsch-Gordan coefficients. Here, $\chi_{\frac{1}{2},m}$ and $Y_{L,m_L}$ denote the spinor of the ground-state baryon and the spherical harmonic function, respectively. The polarization vector for the $\bar{D}_1$ meson is $\epsilon_{m}^{\mu}$ ($m=0,\pm1$), and the polarization tensor for the $\bar{D}_2^*$ meson is constructed as $\zeta_{m}^{\mu\nu}=\sum_{m_1,m_2}\langle1,m_1;1,m_2|2,m\rangle\epsilon_{m_1}^{\mu}\epsilon_{m_2}^{\nu}$ \cite{Cheng:2010yd}, with the explicit representations $\epsilon_{\pm}^{\mu}= (0,\pm1,i,0)/\sqrt{2}$ and $\epsilon_{0}^{\mu}= (0,0,0,-1)$.

\renewcommand\tabcolsep{0.1cm}
\renewcommand{\arraystretch}{1.7}
\begin{table}[!htpb]
\centering
\caption{The flavor wave functions $|I, I_3\rangle$ and spin-orbit wave functions $|{}^{2S+1}L_{J}\rangle$ for the discussed $\bar TB$ systems.}\label{flavor wave}
\begin{tabular}{cl|ccl}\toprule[1.0pt]\midrule[1.0pt]
     &$|I,I_3\rangle$     & &$J^P$   &$|{}^{2S+1}L_J\rangle$ \\\midrule[1.0pt]
$\bar{T}N$ &$|1,1\rangle=\bar{T}^{0}p$
&$\bar{D}_1B$      &$\frac{1}{2}^+$    &$|{}^2\mathbb{S}_{\frac{1}{2}}, {}^4\mathbb{D}_{\frac{1}{2}}\rangle$\\
&$|1,0\rangle=\frac{1}{\sqrt{2}}(\bar{T}^{0}n+{T}^{-}p
)$
  &&$\frac{3}{2}^+$    &$|{}^4\mathbb{S}_{\frac{3}{2}}, {}^2\mathbb{D}_{\frac{3}{2}}, {}^4\mathbb{D}_{\frac{3}{2}}\rangle$\\
               &$|1,-1\rangle={T}^{-}n$ 
    &&$\frac{1}{2}^-$  &$|{}^2\mathbb{P}_{\frac{1}{2}}, {}^4\mathbb{P}_{\frac{1}{2}}\rangle$\\
               &$|0,0\rangle=\frac{1}{\sqrt{2}}(\bar{T}^{0}n-{T}^{-}p)$ 
    &&$\frac{3}{2}^-$ &$|{}^2\mathbb{P}_{\frac{3}{2}}, {}^4\mathbb{P}_{\frac{3}{2}}\rangle$\\

$\bar{{T}} \Lambda$ &$|\frac{1}{2},\frac{1}{2}\rangle=\bar{{T}}^{0}\Lambda^0$
   &&$\frac{5}{2}^-$  &${}^4\mathbb{P}_{\frac{5}{2}}\rangle$\\
&$|\frac{1}{2},-\frac{1}{2}\rangle={T}^{-}\Lambda^0$
     &$\bar{D}_2^*B$  &$\frac{1}{2}^+$   &$|{}^4\mathbb{D}_{\frac{1}{2}},{}^6\mathbb{D}_{\frac{1}{2}}\rangle$\\
     
$\bar{{T}} \Sigma$ &$|\frac{1}{2},\frac{1}{2}\rangle=\sqrt{\frac{1}{3}}\bar{{T}}^{0}\Sigma^0-\sqrt{\frac{2}{3}}{T}^{-}\Sigma^+$
  &&$\frac{3}{2}^+$  &$|{}^4\mathbb{S}_{\frac{3}{2}}, {}^4\mathbb{D}_{\frac{3}{2}}, {}^6\mathbb{D}_{\frac{3}{2}}\rangle$\\
&$|\frac{1}{2},-\frac{1}{2}\rangle=-\sqrt{\frac{1}{3}}{T}^{-}\Sigma^0+\sqrt{\frac{2}{3}}\bar{{T}}^{0}\Sigma^-$ 
   &&$\frac{5}{2}^+$   &$|{}^6\mathbb{S}_{\frac{5}{2}}, {}^4\mathbb{D}_{\frac{5}{2}}, {}^6\mathbb{D}_{\frac{5}{2}}\rangle$\\
 &$|\frac{3}{2},\frac{3}{2}\rangle=\bar{{T}}^{0}\Sigma^+$
   &&$\frac{1}{2}^-$  &$|{}^4\mathbb{P}_{\frac{1}{2}}\rangle$\\
&$|\frac{3}{2},\frac{1}{2}\rangle=\sqrt{\frac{2}{3}}\bar{{T}}^{0}\Sigma^0+\sqrt{\frac{1}{3}}{{T}}^{-}\Sigma^+$ 
   & &$\frac{3}{2}^-$  &$|{}^4\mathbb{P}_{\frac{3}{2}}, {}^6\mathbb{P}_{\frac{3}{2}}\rangle$\\
&$|\frac{3}{2},-\frac{1}{2}\rangle=\sqrt{\frac{2}{3}}{T}^{-}\Sigma^0+\sqrt{\frac{1}{3}}\bar{{T}}^{0}\Sigma^-$ 
    &&$\frac{5}{2}^-$  &$|{}^4\mathbb{P}_{\frac{5}{2}}, {}^6\mathbb{P}_{\frac{5}{2}}\rangle$\\
&$|\frac{3}{2},-\frac{3}{2}\rangle={{T}}^{-}\Sigma^-$
  &&$\frac{7}{2}^-$   &$|{}^6\mathbb{P}_{\frac{7}{2}}\rangle$\\

$\bar{{T}}\Xi$ &$|1,1\rangle=\bar{{T}}^{0}\Xi^0$
\\
&$|1,0\rangle=\frac{1}{\sqrt{2}}(\bar{{T}}^{0}\Xi^-+{{T}}^{-}\Xi^0
)$
\\
               &$|1,-1\rangle={{T}}^{-}\Xi^-$ \\
               &$|0,0\rangle=\frac{1}{\sqrt{2}}(\bar{{T}}^{0}\Xi^--{{T}}^{-}\Xi^0
)$ \\
\bottomrule[1.0pt]\midrule[1.0pt]
\end{tabular}
\end{table}

The OBE effective potentials are derived through the following standard procedure. After constructing the relevant effective Lagrangians, we obtain the scattering amplitudes $\mathcal{M}(\bar{T}_i B_i \to \bar{T}_f B_f)$ for $t$-channel single-meson ($M$) exchange. The momentum-space potential in the nonrelativistic limit is then given by the Breit approximation as
\begin{eqnarray}
  \mathcal{V}_M^{\bar{T}_iB_i\to \bar{T}_fB_f}(q) = -\frac{\mathcal{M}( \bar{T}_iB_i\to \bar{T}_fB_f)}{\sqrt{2m_{\bar{T}_i} 2m_{B_i} 2m_{\bar{T}_f} 2m_{B_f}}}.
  \end{eqnarray}
The contributing mesons $M$ are the scalar $\sigma$, pseudoscalars $\pi$ and $\eta$, and vectors $\rho$ and $\omega$. The coordinate-space potential $\mathcal{V}(r)$ follows from the Fourier transform as
\begin{eqnarray}
\mathcal{V}(r) = \int \frac{d^3\vec{q}}{(2\pi)^{3}}e^{i\vec{q}\cdot r}\mathcal{V}(q)\mathcal{F}^2(q^2,m_E^2).
\end{eqnarray}
Here, a monopole form factor $\mathcal{F}(q^2, m_E^2)=(\Lambda^2 - m_E^2)/(\Lambda^2 - q^2)$ is introduced at each vertex to account for the composite nature of the hadrons and to regulate the high-momentum divergence, where $\Lambda$, $m_E$, and $q$ are the cutoff parameter, mass, and four-momentum of the exchanged meson, respectively. Based on the successful description of the deuteron and other molecular candidates \cite{Tornqvist:1993ng, Tornqvist:1993vu, Chen:2015loa, Liu:2011xc, Chen:2016ypj, Chen:2021vhg, Chen:2020kco, Chen:2020yvq, Chen:2022dad, Chen:2022onm, Wang:2020bjt, Wang:2019nwt, Chen:2019asm, Chen:2019uvv, He:2013nwa, Li:2012ss, Li:2012bt, Sun:2012sy, Sun:2011uh, Yang:2011wz, Lee:2011rka, Liu:2009ei, Liu:2008xz, Chen:2024xlw, Liu:2008fh, Yang:2021sue, Liu:2010xh, Qian:2024joy, Liu:2019tjn, Yamaguchi:2019seo, He:2019ify, Burns:2019iih, He:2011ed, Thomas:2008ja, Liu:2008tn, Lee:2009hy, Wu:2010jy, Huo:2024eew,Cui:2025elw,Tang:2025bcc}, the reasonable cutoffs $\Lambda$ are in the range around 1.00 GeV, as corresponding to the typical hadronic scale or to the intrinsic size of hadrons. As we will see in the following numerical calculations, we attempt to search for bound-state solutions by varying the cutoff parameter in the region of $\Lambda\leq2.00$ GeV to obtain binding energy around a few to 10 MeV for loosely bound molecular hadrons. The physical relevance of the results will be discussed in terms of the cutoff parameters. For  loosely bound systems with the same binding energy, the smaller cutoff parameters correspond to stronger attractive OBE effective potentials, thereby increasing the likelihood of the system being a hadronic molecular candidate.

The effective Lagrangians governing the interactions of the $P$-wave anticharmed meson doublet $(\bar{D}_1, \bar{D}_2^*)$ with light mesons are constructed with heavy-quark symmetry and chiral symmetry \cite{Wise:1992hn, Casalbuoni:1992gi, Casalbuoni:1996pg, Yan:1992gz} as follows:
\begin{eqnarray}\label{eq:lag}
{\cal L}&=
&g_\sigma^{\prime\prime}\left\langle\overline{T}^{\,(\overline{Q})\mu}_b\sigma T^{(\overline{Q})}_{b\mu}\right\rangle+ik\left\langle\overline{T}^{\,(\overline{Q})\mu}_a{\cal A}\!\!\!\slash_{ab}\gamma_5T^{(\overline{Q})}_{b\mu}\right\rangle\nonumber\\
&&-\left\langle i\overline{T}^{\,(\overline{Q})}_{a\lambda}\left(\beta^{\prime\prime} v^{\mu}({\cal V}_{\mu}-\rho_{\mu})-\lambda^{\prime\prime}\sigma^{\mu\nu}F_{\mu\nu}(\rho)\right)_{ab}T^{(\overline{Q})\lambda}_{b}\right\rangle,~~
\end{eqnarray}
where the axial current ${\cal A}_{\mu}=\frac{1}{2}\left(\xi^{\dagger}\partial_{\mu}\xi-\xi\partial_{\mu}\xi^{\dagger}\right)_{\mu}$, vector current ${\cal V}{\mu}=\frac{1}{2}\left(\xi^{\dagger}\partial_{\mu}\xi+\xi\partial_{\mu}\xi^{\dagger}\right)_{\mu}$, and vector meson field strength $F_{\mu\nu}(\rho)=\partial_{\mu}\rho_{\nu}-\partial_{\nu}\rho_{\mu}+[\rho_{\mu},\rho_{\nu}]$ are defined in the standard chiral perturbation theory formalism. The pseudo-Goldstone and vector meson matrices are $\xi=\exp(i\mathbb{P}/f_{\pi})$ and $\rho_{\mu}=ig_V\mathbb{V}_{\mu}/\sqrt{2}$, respectively. The superfield $T^{(\overline{Q})\mu}{a}$, which combines the $\bar{D}_1$ and $\bar{D}_2^*$ states, is
\begin{eqnarray}
T^{(\overline{Q})\mu}_{a}&=&\left[\bar{D}^{*(\overline{Q})\mu\nu}_{2a}\gamma_{\nu}-\sqrt{\frac{3}{2}}\bar{D}^{(\overline{Q})}_{1a\nu}\gamma_5\left(g^{\mu\nu}-\frac{1}{3}\left(\gamma^{\mu}-v^{\mu}\right)\gamma^{\nu}\right)\right]{\cal P}_{-}.\nonumber
\end{eqnarray}
Here, ${\cal P}_{-}=(1-{v}\!\!\!\slash)/2$ is the heavy-quark projection operator, and $v^{\mu}=(1,0,0,0)$ is the four-velocity of the heavy quark in the nonrelativistic approximation.

After expanding Eq. (\ref{eq:lag}), we further obtain the explicit interaction Lagrangians as
\begin{eqnarray}
\label{eq:lagp}
\mathcal{L}_{\bar{D}_1 \bar{D}_1\sigma} &=& -2g_\sigma^{\prime\prime}\bar{D}_{1a\mu}\bar{D}^{\mu\dagger}_{1a} \sigma ,\\
\mathcal{L}_{\bar{D}^*_2\bar{D}^*_2\sigma} &=& 2g_\sigma^{\prime\prime}\bar{D}^{*\dagger}_{2a\mu\nu}\bar{D}^{*\mu\nu}_{2a}\sigma ,\\
\mathcal{L}_{\bar{D}_1\bar{D}^*_2\sigma} &=& \sqrt{\frac{2}{3}}ig_\sigma^{\prime\prime}\epsilon^{\mu \nu \rho \tau}v_{\rho}(\bar{D}^{\dagger}_{1a\nu}\bar{D}^{*}_{2a\mu\tau}-\bar{D}_{1a\nu}\bar{D}^{*\dagger}_{2a\mu\tau}) \sigma ,\\
\mathcal {L}_{\bar{D}_1 \bar{D}_1\mathbb{P}}&=&-\frac{5ik}{3f_\pi}~\varepsilon^{\mu\nu\rho\tau}v_\nu
    \bar{D}^{\dagger}_{1a\rho}\bar{D}_{1b\tau} \partial_\mu\mathbb{P}_{ab},\\
\mathcal {L}_{\bar{D}^*_2\bar{D}^*_2\mathbb{P}}&=&\frac{2ik}{f_\pi}~\varepsilon^{\mu\nu\rho\tau}v_\nu
   \bar{D}^{*\alpha\dagger}_{2a\rho}\bar{D}^{*}_{2b\alpha\tau}\partial_\mu\mathbb{P}_{ab},\\
\mathcal {L}_{\bar{D}_1\bar{D}^*_2\mathbb{P}}&=&\sqrt{\frac{2}{3}}\frac{k}{f_\pi}(
\bar{D}^{*\mu\lambda}_{2a}\bar{D}^{\dagger}_{1b\mu}+\bar{D}_{1a\mu}\bar{D}^{*\mu\lambda\dagger}_{2b})
\partial_\lambda\mathbb{P}_{ab},\\
\mathcal {L}_{\bar{D}_1 \bar{D}_1\mathbb{V}} &=& \sqrt{2}\beta^{\prime \prime}
g_{V}(v\cdot\mathbb{V}_{ab}) \bar{D}_{1a\mu}\bar{D}^{\mu\dagger}_{1b}\nonumber\\
&&+\frac{5\sqrt{2}i\lambda^{\prime\prime} g_{V}}{3}(\bar{D}^{\nu}_{1a}\bar{D}^{\mu\dagger}_{1b}-\bar{D}^{\nu\dagger}_{1a}\bar{D}^{\mu}_{1b})\partial_\mu \mathbb{V}_{ab\nu},\\
\mathcal {L}_{\bar{D}^*_2\bar{D}^*_2\mathbb{V}} &=& -\sqrt{2}\beta^{\prime \prime}
g_{V}(v\cdot\mathbb{V}_{ab}) \bar{D}_{2a}^{*\lambda\nu}  \bar{D}^{*\dagger}_{2b{\lambda\nu}}+2\sqrt{2}i\lambda^{\prime\prime} g_{V}\nonumber\\
&&\times(\bar{D}^{*\lambda\nu\dagger}_{2a} \bar{D}^{*\mu}_{2b\lambda}-\bar{D}^{*\lambda\nu}_{2a}\bar{D}^{*\mu\dagger}_{2b\lambda} )\partial_\mu \mathbb{V}_{ab\nu},\\
\mathcal {L}_{\bar{D}_1\bar{D}^*_2\mathbb{V}} &=& \frac{i\beta^{\prime \prime}g_{V}}{\sqrt{3}}\varepsilon^{\lambda\alpha\rho\tau}v_{\rho}(v\cdot
\mathbb{V}_{ab})(\bar{D}^{\dagger}_{1a\alpha}\bar{D}^{*}_{2b\lambda\tau}-\bar{D}_{1a\alpha}
\bar{D}^{\dagger*}_{2b\lambda\tau})\nonumber\\
&&+\frac{2\lambda^{\prime\prime} g_{V}}{\sqrt{3}}[3\varepsilon^{\mu\lambda\nu\tau}v_\lambda(\bar{D}^{\alpha\dagger}_{1a}
\bar{D}^{*}_{2b\alpha\tau}+\bar{D}^{\alpha}_{1a}\bar{D}^{*\dagger}_{2b\alpha\tau})\partial_\mu \mathbb{V}_{ab\nu}\nonumber\\
&&+2\varepsilon^{\lambda\alpha\rho\nu}v_\rho(\bar{D}^{\dagger}_{1a\alpha}\bar{D}^{*\mu}_{2b\lambda}
+\bar{D}_{1a\alpha}\bar{D}^{\dagger\mu*}_{2b\lambda})\nonumber\\
&&\times(\partial_\mu \mathbb{V}_{ab\nu}-\partial_\nu \mathbb{V}_{ab\mu})].
\end{eqnarray}

As for the interactions between the light octet baryons $(B)$ and light mesons, we employ the following $SU(3)$-symmetric Lagrangians as
\begin{eqnarray}
\mathcal{L}_{B B \sigma} &  = & g_{B B \sigma} \bar{B} \sigma B, \\
\mathcal{L}_{B B \mathbb{P}} & = & \sqrt{2} \frac{g_{B B P}}{m_{P}} \bar{B} \gamma^{5} \gamma^{\mu} \partial_{\mu} \mathbb{P}_{ba} B, \\
\mathcal{L}_{B B \mathbb{V}} & = & \sqrt{2} g_{B B V} \bar{B}\gamma^{\mu}\mathbb{V}_{ba\mu}B-\frac{f_{B B V}}{\sqrt{2} m_{B}}\bar{B} \sigma^{\mu v} \partial_{v} \mathbb{V}_{ba\mu} B,
\end{eqnarray}
where pseudoscalar meson matrices $\mathbb{P}$ and vector meson matrices $\mathbb{V}_{\mu}$ are expressed as
\begin{eqnarray*}
\mathbb{P}&=& \left(\begin{array}{ccc}
\frac{\pi^0}{\sqrt{2}}+\frac{\eta}{\sqrt{6}} &\pi^+ &K^+ \\
\pi^- &-\frac{\pi^0}{\sqrt{2}}+\frac{\eta}{\sqrt{6}} &K^0 \\
K^- &\bar{K}^0 &-\sqrt{\frac{2}{3}}\eta
\end{array}\right),\\
\mathbb{V}_{\mu}&=& \left(\begin{array}{ccc}
\frac{\rho^0}{\sqrt{2}}+\frac{\omega}{\sqrt{2}} &\rho^+ &K^{*+} \nonumber\\
\rho^- &-\frac{\rho^0}{\sqrt{2}}+\frac{\omega}{\sqrt{2}} &K^{*0} \nonumber\\
K^{*-} &\bar{K}^{*0} &\phi
\end{array}\right)_{\mu},\\
B &=& \left(\begin{array}{ccc}
\frac{\Sigma^0}{\sqrt{2}}+\frac{\Lambda}{\sqrt{6}}    &\Sigma^+    &p\\
\Sigma^-  &-\frac{\Sigma^0}{\sqrt{2}}+\frac{\Lambda}{\sqrt{6}}   &n\\
\Xi^-      &\Xi^0     &-\frac{2}{\sqrt{6}}\Lambda\end{array}\right).
\end{eqnarray*}
respectively. Then, for all the coupling constants, they are estimated from well-established nucleon-nucleon interactions through the quark model and $SU(3)$ symmetry relations \cite{Wang:2019aoc, Wang:2011rga, Machleidt:2000ge, Machleidt:1987hj, Cao:2010km}, whose numerical values are compiled in Table \ref{constants}.

\renewcommand\tabcolsep{0.12cm}
\renewcommand{\arraystretch}{1.7}
\begin{table}[!htbp]
\caption{Coupling constants adopted in our calculations.}\label{constants}
{\begin{tabular}{cccccc}
\toprule[1pt]
\toprule[1pt]
$\frac{g^2_{\sigma NN}}{4\pi}=5.69$   &$g_{NN\eta}=0.33$   &$\frac{g^2_{\pi NN}}{4\pi}=0.07$   &$g_{\Lambda\Lambda\omega}=7.98$\\   
$\frac{g^2_{\rho NN}}{4\pi}=0.81$ &$f_{\Lambda\Lambda\omega}=-9.73$
&$\frac{f_{\rho NN}}{g_{\rho NN}}=6.10$ &$g_{\Lambda\Lambda\eta}=-0.67$\\
$\frac{g_{\omega NN}^2}{4\pi}=20.00$ 
&$\frac{f_{\omega NN}}{g_{\omega NN}}=0.00$   &$k=0.78$   &$f_\pi=0.13$\\
$f_{\Lambda\Sigma\rho}=16.85$  &$g_{\Lambda\Sigma\rho}=-0.55$
&$g_{\Lambda\Sigma\pi}=0.67$   &$g_{N \Sigma K}=0.19$\\
$g_{ \Sigma\Sigma \rho/\omega}=7.34$  &$f_{\Sigma\Sigma \rho/\omega}=9.73$
&$g_{\Sigma\Sigma \pi}=0.77$   &$g_{\Sigma\Sigma \eta}=0.67$\\
$g_{\Xi\Xi\pi}=-0.19$     &$g_{\Xi\Xi\eta}=-0.99$
&$g_{\Xi\Xi\rho/\omega}=4.23$   &$\lambda^{\prime\prime} g_V=7.38$    \\
$f_{\Xi\Xi\rho/\omega}=-9.91$        &$g_\sigma=2.82$
&$\beta^{\prime\prime} g_V=-6.50$
 \\     
\bottomrule[1pt]\bottomrule[1pt]
\end{tabular}}
\end{table}

\renewcommand\tabcolsep{0.25cm}
\renewcommand{\arraystretch}{1.8}
\begin{table}[!htpb]
\centering
\caption{The isospin factor for diﬀerent exchanged mesons in our calculations.} \label{isospin}
\begin{tabular}{ccccccc}\toprule[1.0pt]\midrule[1.0pt]
     &$I$     &$\mathcal{G}_{\sigma}$    & $\mathcal{G}_{\pi}$    &$\mathcal{G}_{\eta}$    &$\mathcal{G}_{\rho}$    &$\mathcal{G}_{\omega}$\\\midrule[1.0pt]
$\bar{T}N-\bar{T}N$  &0    &1   &$-\frac{3}{2}$   &$\frac{\sqrt{3}}{6}$   &$-\frac{3}{2}$   &$\frac{1}{2}$   
\\
&1    &1   &$\frac{1}{2}$   &$\frac{\sqrt{3}}{6}$   &$\frac{1}{2}$   &$\frac{1}{2}$   
\\
$\bar{T}\Lambda-\bar{T}\Lambda$  &$\frac{1}{2}$    &1   &$\cdots$   &$\frac{\sqrt{3}}{6}$   &$\cdots$   &$\frac{1}{2}$   
\\
$\bar{T}\Sigma-\bar{T}\Sigma$  &$\frac{1}{2}$    &1   &$-1$   &$\frac{\sqrt{3}}{6}$   &$-1$   &$\frac{1}{2}$   
\\
&$\frac{3}{2}$    &1   &$\frac{1}{2}$   &$\frac{\sqrt{3}}{6}$   &$\frac{1}{2}$   &$\frac{1}{2}$   
\\
$\bar{T}\Xi-\bar{T}\Xi$  &0    &1   &$-\frac{3}{2}$   &$\frac{1}{2\sqrt{3}}$   &$-\frac{3}{2}$   &$\frac{1}{2}$   
\\
&1    &1   &$\frac{1}{2}$   &$\frac{1}{2\sqrt{3}}$   &$\frac{1}{2}$   &$\frac{1}{2}$   
\\
$\bar{T}\Lambda-\bar{T}\Sigma$  &$\frac{1}{2}$    &$\cdots$   &$\frac{\sqrt{3}}{2}$   &$\cdots$   &$\frac{\sqrt{3}}{2}$   &$\cdots$   
\\
\bottomrule[1.0pt]\midrule[1.0pt]
\end{tabular}
\end{table}

\renewcommand\tabcolsep{0.25cm}
\renewcommand{\arraystretch}{2}
\begin{table*}[!htbp]
\caption{Matrix elements for the spin-spin interactions and tensor force operators in the OBE effective potentials. Here, $\langle\mathcal{A}_{1}\rangle=\langle\mathcal{A}_{4}\rangle=\mathcal{I}$ with $\mathcal{I}$ being the identity matrix.}\label{operator_transposed}
{\begin{tabular}{ccccccc}
\toprule[1pt]
\toprule[1pt]
$J^P$ & $\mathcal{A}_{2}$ & $\mathcal{A}_{3}$ & $\mathcal{A}_{7}$ & $\mathcal{A}_{8}$ & $\mathcal{A}_{5}$ & $\mathcal{A}_{6}$ \\\hline
\addlinespace[0.3cm]
$\frac{1}{2}^-$ &
$\begin{pmatrix}-2 & 0\\ 0 & 1 \end{pmatrix}$ &
$\begin{pmatrix}0 & -\sqrt{2}\\ -\sqrt{2} & -2 \end{pmatrix}$ &
$\begin{pmatrix}0 \\ \sqrt{\frac{5}{2}} \end{pmatrix}$ &
$\begin{pmatrix}\sqrt{\frac{1}{5}} \\ \sqrt{\frac{2}{5}} \end{pmatrix}$ &
$\begin{pmatrix}-\frac{3}{2} & 0\\ 0 & 1 \end{pmatrix}$ &
$\begin{pmatrix}-\frac{12}{25} & -\frac{63}{50}\\ -\frac{63}{50} & -\frac{28}{25} \end{pmatrix}$ \\
\addlinespace[0.3cm]
$\frac{3}{2}^-$ &
$\begin{pmatrix}-2 & 0\\ 0 & 1 \end{pmatrix}$ &
$\begin{pmatrix}0 & \frac{1}{\sqrt{5}}\\ \frac{1}{\sqrt{5}} & \frac{8}{5} \end{pmatrix}$ &
$\begin{pmatrix}0 & 0\\ \sqrt{\frac{5}{2}} & 0 \end{pmatrix}$ &
$\begin{pmatrix}-\frac{1}{5\sqrt{2}} & -\frac{6\sqrt{2}}{5}\\ -\frac{4}{5}\sqrt{\frac{2}{5}} & -\frac{21}{5\sqrt{10}} \end{pmatrix}$ &
$\begin{pmatrix}-\frac{3}{2} & 0\\ 0 & 1 \end{pmatrix}$ &
$\begin{pmatrix}\frac{3}{25} & \frac{9}{25}\sqrt{\frac{7}{2}}\\ \frac{9}{25}\sqrt{\frac{7}{2}} & \frac{32}{25} \end{pmatrix}$ \\
\addlinespace[0.3cm]
$\frac{5}{2}^-$ &
$\begin{pmatrix}-1 \end{pmatrix}$ &
$\begin{pmatrix}0 \end{pmatrix}$ &
$\begin{pmatrix}0 & \sqrt{\frac{5}{2}} \end{pmatrix}$ &
$\begin{pmatrix}\frac{1}{5}\sqrt{\frac{2}{5}} & \frac{3}{5}\sqrt{\frac{7}{5}} \end{pmatrix}$ &
$\begin{pmatrix}0 \end{pmatrix}$ &
$\begin{pmatrix}\frac{3}{5} \end{pmatrix}$\\\addlinespace[0.3cm]
$\frac{1}{2}^+$ &
$\begin{pmatrix}-2 & 0\\ 0 & 1 \end{pmatrix}$ &
$\begin{pmatrix}0 & -\sqrt{2}\\ -\sqrt{2} & -2 \end{pmatrix}$ &
$\begin{pmatrix}0 & 0\\ \sqrt{\frac{5}{2}} & 0 \end{pmatrix}$ &
$\begin{pmatrix}\frac{1}{\sqrt{5}} & 2\sqrt{\frac{6}{5}}\\ \sqrt{\frac{2}{5}} & -\sqrt{\frac{3}{5}} \end{pmatrix}$ &
$\begin{pmatrix}-\frac{3}{2} & 0\\ 0 & 1 \end{pmatrix}$ &
$\begin{pmatrix}\frac{3}{5} & -\frac{3}{5}\sqrt{\frac{3}{2}}\\ -\frac{3}{5}\sqrt{\frac{3}{2}} & -\frac{8}{5} \end{pmatrix}$ \\
\addlinespace[0.3cm]
$\frac{3}{2}^+$ &
$\begin{pmatrix}1 & 0 & 0\\ 0 & -2 & 0\\ 0 & 0 & 1 \end{pmatrix}$ &
$\begin{pmatrix}0 & 1 & 2\\ 1 & 0 & -1\\ 2 & -1 & 0 \end{pmatrix}$ &
$\begin{pmatrix}\sqrt{\frac{5}{2}} & 0 & 0\\ 0 & 0 & 0\\ 0 & \sqrt{\frac{5}{2}} & 0 \end{pmatrix}$ &
$\begin{pmatrix}0 & -\sqrt{\frac{2}{5}} & -\sqrt{\frac{21}{10}}\\ -\frac{1}{\sqrt{10}} & \frac{1}{\sqrt{10}} & -2\sqrt{\frac{6}{35}}\\ -\sqrt{\frac{2}{5}} & 0 & -\sqrt{\frac{15}{14}} \end{pmatrix}$ &
$\begin{pmatrix}-\frac{3}{2} & 0 & 0\\ 0 & -\frac{3}{2} & 0\\ 0 & 0 & 1 \end{pmatrix}$ &
$\begin{pmatrix}0 & -\frac{3}{5} & -\frac{3\sqrt{21}}{10}\\ -\frac{3}{5} & 0 & -\frac{3}{2}\sqrt{\frac{3}{7}}\\ -\frac{3\sqrt{21}}{10} & -\frac{3}{2}\sqrt{\frac{3}{7}} & -\frac{4}{7} \end{pmatrix}$ \\
\addlinespace[0.3cm]
$\frac{5}{2}^+$ &
$\cdots$ &
$\cdots$ &
$\cdots$ &
$\cdots$ &
$\begin{pmatrix}1 & 0 & 0\\ 0 & -\frac{3}{2} & 0\\ 0 & 0 & 1 \end{pmatrix}$ &
$\begin{pmatrix}0 & \frac{3}{5}\sqrt{\frac{7}{2}} & \frac{2\sqrt{14}}{5}\\ \frac{3}{5}\sqrt{\frac{7}{2}} & -\frac{3}{7} & -\frac{3}{7}\\ \frac{2\sqrt{14}}{5} & -\frac{3}{7} & \frac{4}{7} \end{pmatrix}$ \\
$\frac{7}{2}^-$ &
$\cdots$ &
$\cdots$ &
$\cdots$ &
$\cdots$  &
$\begin{pmatrix}1 \end{pmatrix}$ &
$\begin{pmatrix}-\frac{2}{5} \end{pmatrix}$\\
\bottomrule[1pt]
\bottomrule[1pt]
\end{tabular}}
\end{table*}

The total OBE effective potential for a given $\bar{T}B$ channel is a sum over the exchanged mesons $M$ as
\begin{eqnarray}
    V_{\bar{T}B} &=& \sum_M \mathcal{G}_{M} \mathcal{V}_{\bar{T}B\to\bar{T}B}^M,
\end{eqnarray}
where $\mathcal{G}_M$ is the isospin factor listed in Table \ref{isospin}, and $\mathcal{V}_{\bar{T}B\to\bar{T}B}^M$ are the OBE subpotentials for the process $\bar{T}B\to\bar{T}
B$ by exchanging meson $M$, i.e., 
\begin{eqnarray}
\mathcal{V}_{\bar{D}_1B\to\bar{D}_1B}^{\sigma} &=&-C_1\mathcal{A}_1 Y_{\Lambda,m_\sigma},\\
\mathcal{V}_{\bar{D}_1B\to\bar{D}_1B}^{\pi,\eta} &=&
\frac{5C_2}{9\sqrt{2}}(\mathcal{A}_2 Z_{\Lambda,m_P} 
 +\mathcal{A}_3 T_{\Lambda,m_P}), 
  \label{pot1}\\
\mathcal{V}_{\bar{D}_1B\to\bar{D}_1B}^{\rho,\omega} 
  &=&-\frac{5C_3}{9}\mathcal{A}_2 Z_{\Lambda,m_V}+\frac{5C_3}{18}\mathcal{A}_3 T_{\Lambda,m_V}\nonumber\\
  &&
  -C_4 \mathcal{A}_1 Y_{\Lambda,m_V},\\
  \mathcal{V}_{\bar{D}_2^*B\to\bar{D}_2^*B}^{\sigma} &=&-C_1 \mathcal{A}_4 Y_{\Lambda,m_\sigma},\\
\mathcal{V}_{\bar{D}_2^*B\to\bar{D}_2^*B}^{\pi,\eta} &=& \frac{\sqrt2 C_2}{3}\left(\mathcal{A}_5 Z_{\Lambda,m_P}+\mathcal{A}_6 T_{\Lambda,m_P}\right),\\
\mathcal{V}_{\bar{D}_2^*B\to\bar{D}_2^*B}^{\rho/\omega}&=&\frac{C_3}{3}\mathcal{A}_6T_{\Lambda,m_V}-\frac{2C_3}{3} \mathcal{A}_5 Z_{\Lambda,m_V}\nonumber\\
&&-C_4 \mathcal{A}_4 Y_{\Lambda,m_V},
\\
\mathcal{V}_{\bar{D}_1B\to\bar{D}_2^*B}^{\pi} &=&
-\frac{ C_2}{3\sqrt{3}}(\mathcal{A}_7Z_{\Lambda,m_P}+\mathcal{A}_8 T_{\Lambda,m_P}),\\
\mathcal{V}_{\bar{D}_1B\to\bar{D}_2^*B}^{\rho} &=&\frac{\sqrt{6}C_3}{18}(2\mathcal{A}_7Z_{\Lambda,m_P}+\mathcal{A}_8T_{\Lambda,m_P}).
\end{eqnarray}
Here, we have defined several coupling constants, operators, and useful functions, whose explicit forms are
\begin{align}
C_1 &= g_{\sigma}g_{\sigma BB}, \quad 
C_2 = \frac{g_{BBP}k}{f_\pi m_{\pi,\eta}}, \quad
C_4 = g_{BBV}\beta^{\prime \prime}g_v,\\
C_3 &= \frac{5\lambda^{\prime \prime}g_v }{9} \left( \frac{g_{BBV}}{2m_2}+\frac{g_{BBV}}{2m_4}+\frac{f_{BBV}}{m_B} \right),\\
\mathcal{A}_1&=\chi^{\dagger}_4 \chi^{\dagger}_3 \bm{\epsilon}^{\dagger}_3\cdot \bm{\epsilon}_1 \chi_2 \chi_1,\label{operator1}\\
\mathcal{A}_2&=\chi_4^\dagger\chi_3^\dagger(i\bm\epsilon_1\times\bm\epsilon_3^\dagger\cdot\bm\sigma)\chi_2\chi_1,\\
\mathcal{A}_3&=\chi_4^\dagger\chi_3^\dagger S({\bm i{\bm\epsilon_1}
\times{\bm\epsilon_3^\dagger},\bm\sigma},\hat{\bm{r}})\chi_2\chi_1,\\
\mathcal{A}_4&=\sum_{m,n,a,b}C^{2,m+n}_{1m,1n}C^{2,a+b}_{1a,1b}\chi_4^\dagger\chi_3^\dagger(\epsilon_1^a\cdot\epsilon_3^{m\dagger})(\epsilon_1^b\cdot\epsilon_3^{n\dagger})\chi_2\chi_1,\\
\mathcal{A}_5&=\sum_{m,n,a,b}C^{2,m+n}_{1m,1n}C^{2,a+b}_{1a,1b}\chi_4^\dagger\chi_3^\dagger(\bm\epsilon_1^a\cdot\bm\epsilon_3^{m\dagger})(i\bm\epsilon_1^b\times\bm\epsilon_3^{n\dagger})\cdot\bm\sigma \chi_2\chi_1,\\
\mathcal{A}_6&=\sum_{m,n,a,b}C^{2,m+n}_{1m,1n}C^{2,a+b}_{1a,1b}\chi_4^\dagger\chi_3^\dagger(\bm\epsilon_1^a\cdot\bm\epsilon_3^{m\dagger})\nonumber\\
  &\quad\quad\times S({i\bm\epsilon_1^b\times\bm\epsilon_3^{n\dagger}},\bm\sigma,\hat{\bm{r}})\chi_2\chi_1,\\
\mathcal{A}_7&=\sum_{m,n,a,b}C^{2,m+n}_{1m,1n}\chi_4^\dagger\chi_3^\dagger(\bm\epsilon_1\cdot\bm\epsilon_3^{m\dagger})(\bm\epsilon_3^{n\dagger}\cdot\bm\sigma) \chi_2\chi_1,\\
\mathcal{A}_8&=\sum_{m,n,a,b}C^{2,m+n}_{1m,1n}\chi_4^\dagger\chi_3^\dagger(\bm\epsilon_1\cdot\bm\epsilon_3^{m\dagger})S({\bm\epsilon_3^{n\dagger}},\bm\sigma,\hat{\bm{r}})\chi_2\chi_1,\label{operator2}\\
Y_{\Lambda,m}&=\frac{1}{4\pi r}(e^{-mr}-e^{-\Lambda r})-\frac{\Lambda^2-m^2}{8 \pi\Lambda}e^{-\Lambda r},\\
T_{\Lambda,m}&=r\frac{\partial}{\partial r}\frac{1}{r}\frac{\partial}{\partial r}Y_{\Lambda,m},\quad\quad
Z_{\Lambda,m}=\nabla^2Y_{\Lambda,m}.
\end{align}
When performing the numerical calculations, the operators in Eqs. (\ref{operator1})$-$(\ref{operator2}) will be replaced by numerical matrices, which are summarized in Table \ref{operator_transposed}.

\section{Numerical results}\label{sec3}

Following the derivation of the OBE effective potentials, we investigate the possible existence of anticharm molecular pentaquarks with strangeness $|S|=0, 1,$ and $2$. We numerically solve the coupled-channel Schr\"{o}dinger equations to search for bound-state solutions, and the identification criteria on the loosely bound molecules are a binding energy $E$ on the order of several to several tens of MeV and a root-mean-square (rms) radius $r_{\text{rms}}$ of approximately 1.00 fm or larger. In our analysis, the cutoff parameter $\Lambda$ is varied within the range $\Lambda \leq 2.00$ GeV. Drawing upon the established phenomenology of nucleon-nucleon interactions \cite{Tornqvist:1993ng, Tornqvist:1993vu}, we regard loosely bound states that emerge with a cutoff $\Lambda \sim 1.00$ GeV as the most promising candidates for anticharm molecular pentaquarks.

\subsection{The $\bar{D}_1N$ and $\bar{D}_2^*N$ systems}\label{secd1n}

\begin{figure*}[!htpb]
\centering
\includegraphics[width=6.5in]{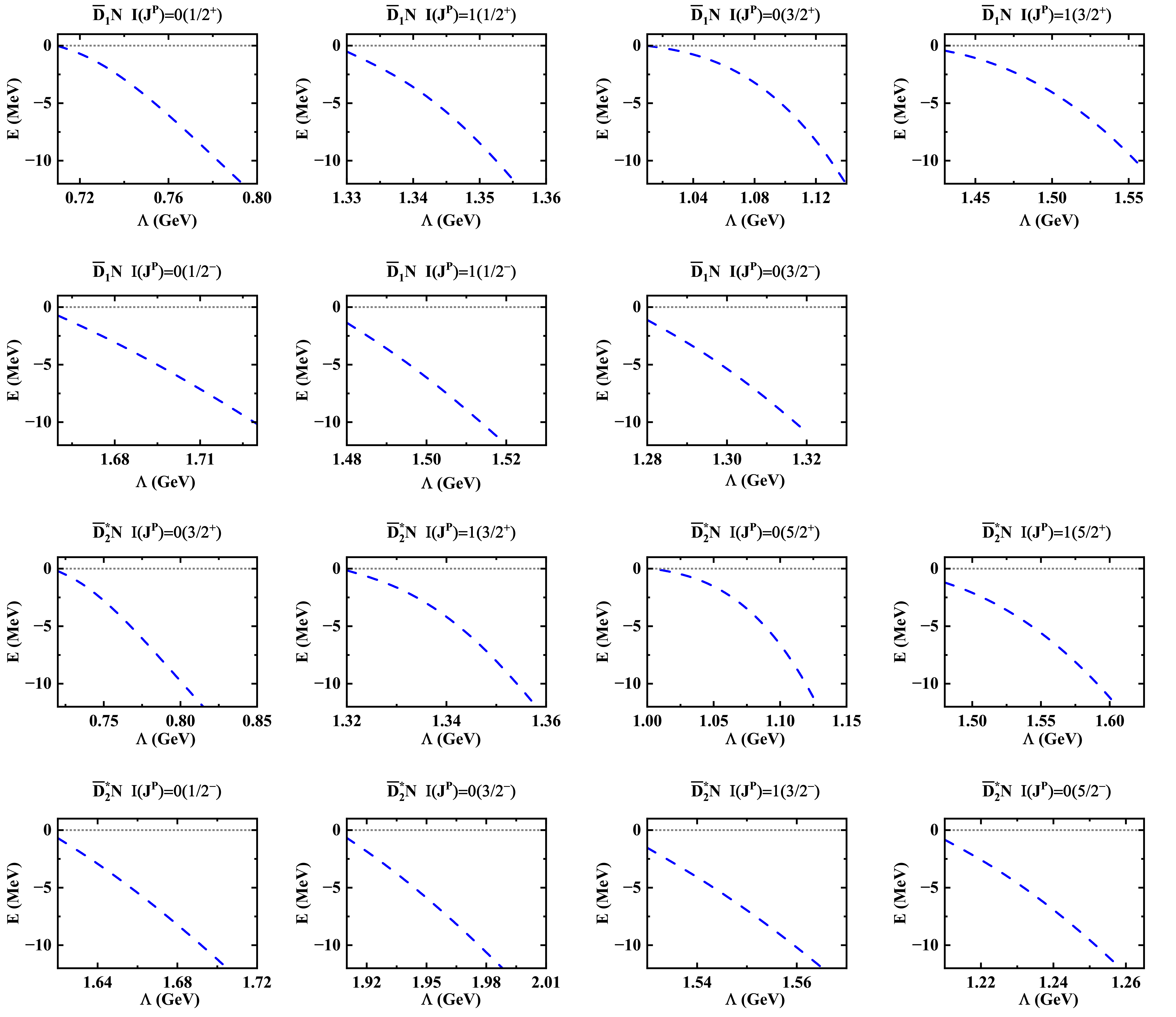}

\caption{The $\Lambda$ dependence of the binding energies $E$ for the ${\bar{T}N}$ bound states.}
\label{TN}
\end{figure*}

\renewcommand\tabcolsep{0.32cm}
\renewcommand{\arraystretch}{1.7}
\begin{table*}[!htpb]
\centering
\caption{The bound-state solutions for the $\bar{T}N$ systems. Here, the units for the cutoff $\Lambda$, the binding energy $E$, and the root-mean-square radius $r_{rms}$ are GeV, MeV, and femtometer, respectively. Channels that are forbidden or dominant are marked with the symbols “$\cdots$” and in bold manner, respectively. The mass thresholds for the involved channels are $m_{\bar{D}_1N}=3360.27$ and $m_{\bar{D}_2^*N}=3401.32$ MeV, respectively.}\label{d1n}

\begin{tabular}{ccccl|ccccl}
\toprule[1.5pt]
\toprule[1.5pt]
$\quad I(J^P)\quad$  &$\Lambda$    &$E$    &$r_{rms}$   &$\bar{D}_1N(^{2}S/^{4}D)$ &$\quad I(J^P)\quad$  &$\Lambda$    &$E$    &$r_{rms}$    &$\bar{D}_2^*N(^{4}S/^{4}D/^{6}D)$\\ 
 $0(\frac{1}{2}^+)$  &0.78   &$-$9.57&1.51&\textbf{99.09} /0.91              &$0(\frac{3}{2}^+)$  &0.78   &$-$6.77&1.76&\textbf{98.32} /0.23/1.45   \\
     &0.79     &$-$11.39    &  1.41  &\textbf{99.08}/0.92     
     &&0.79     &$-$8.26    & 1.63  &\textbf{98.27}/0.25/1.48\\
 $1(\frac{1}{2}^+)$  &1.33   &$-$0.51&4.72&\textbf{97.80}/2.20             &$1(\frac{3}{2}^+)$  &1.33   &$-$1.42&3.29&\textbf{96.03}/0.77/3.20      \\
     &1.35     &$-$8.13    & 1.40  &\textbf{94.75} /5.25    &&1.34     &$-$3.95    &  2.04  &\textbf{94.36} /1.11/4.53 \\\hline
   $I(J^P)$  &$\Lambda$    &$E$    &$r_{rms}$      &$\bar{D}_1N(^{4}S/^{2}D/^{4}D)$ &$\quad I(J^P)\quad$   &$\Lambda$    &$E$    &$r_{rms}$   &$\bar{D}_2^*N(^{6}S/^{4}D/^{6}D)$ \\
$0(\frac{3}{2}^+)$   & 1.05  &$-$1.17  &4.00  &\textbf{95.52}/0.65/3.83                 & $0(\frac{5}{2}^+)$   & 1.05  &$-$1.52  &3.63  &\textbf{94.30}/1.62/4.08      \\
            & 1.10  &$-$5.31  &2.16  &\textbf{92.15}/1.11/6.74
          & & 1.10  &$-$6.55  &2.00  &\textbf{90.22}/2.75/7.03               \\ 
            $1(\frac{3}{2}^+)$   &1.45  &$-$1.06  &3.93  &\textbf{95.64}/1.03/3.33    
           & $1(\frac{5}{2}^+)$   &1.50  &$-$2.08  &3.00  &\textbf{94.75}/2.21/3.04          \\
           &  1.50  &$-$4.03  &2.26  &\textbf{92.00}/1.95/6.05
               & &1.55  &$-$5.57  &1.99  &\textbf{91.88}/3.48/4.64           \\\midrule[1.5pt]
$I(J^P)$  &$\Lambda$    &$E$    &$r_{rms}$     &$\bar{D}_1N(^{2}P/^{4}P)$   & $\quad I(J^P)\quad$   &$\Lambda$    &$E$    &$r_{rms}$   &$\bar{D}_2^*N(^{4}P/^{6}P)$\\ 
 $0(\frac{1}{2}^-)$    &1.66    &$-$0.72    &2.39    &\textbf{83.00}/17.00      
&$0(\frac{1}{2}^-)$   &1.63  &$-$1.75  &1.98  &\textbf{100.00}   /$\cdots$ 
   \\
 &1.70    &$-$5.68    &1.49    &\textbf{83.55}/16.45      
  &&1.66  &$-$5.43  &1.52  & \textbf{100.00}   /$\cdots$
   \\ 
 $1(\frac{1}{2}^-)$& 1.48     &$-$1.35    &2.14    &17.57/\textbf{82.43}      
   &$0(\frac{3}{2}^-)$   &1.91  &$-$0.67  &2.33  &\textbf{69.68}  /30.32 \\
&1.65     &$-$7.27    &1.39    &17.81/\textbf{82.19}      
 & &1.96  &$-$7.37  &1.33  &\textbf{70.40}   /29.60 
   \\ 
  $0(\frac{3}{2}^-)$     &1.28      &$-$1.11    &2.50     &10.89 /\textbf{89.11}      
  & $1(\frac{3}{2}^-)$   &1.53  &$-$1.52  &2.03  &32.14   /\textbf{67.86} 
   \\
 &1.3      &$-$5.30    &1.68     &11.27 /\textbf{88.73}     & &1.55  &$-$6.90  &1.39  &32.97   /\textbf{67.03 }
    \\
  $1(\frac{3}{2}^-)$     &$\cdots$      &$\cdots$    &$\cdots$     &$\cdots$ /$\cdots$       
&$0(\frac{5}{2}^-)$   &1.21  &$-$0.83  &2.75  &25.15   /\textbf{74.85} 
  \\
&$\cdots$      &$\cdots$    &$\cdots$     &$\cdots$ /$\cdots$       
 &&1.24  &$-$6.87  &1.61  &26.46   /\textbf{73.54} 
  \\
    
$0(\frac{5}{2}^-)$     &$\cdots$      &$\cdots$    &$\cdots$     &$\cdots$ /$\cdots$      
  &$1(\frac{1}{2}^-)$   &$\cdots$  &$\cdots$  &$\cdots$  &$\cdots$   /$\cdots$ 
  \\
 $1(\frac{5}{2}^-)$     &$\cdots$      &$\cdots$    &$\cdots$     &$\cdots$ /$\cdots$       
  &$1(\frac{5}{2}^-)$    &$\cdots$  &$\cdots$  &$\cdots$  &$\cdots$   /$\cdots$ \\
  &      &    &     &       
  &$0(\frac{7}{2}^-)$    &$\cdots$  &$\cdots$  &$\cdots$  &$\cdots$   /$\cdots$ \\
     &      &    &     &       
  &$1(\frac{7}{2}^-)$    &$\cdots$  &$\cdots$  &$\cdots$  &$\cdots$   /$\cdots$ \\
\bottomrule[1.5pt]
\bottomrule[1.5pt]
\end{tabular}
\end{table*}

The bound-state solutions for the $\bar{D}_1N$ and $\bar{D}_2^*N$ systems, incorporating $S$-$D$ wave mixing and $P$-wave interactions, are summarized in Table \ref{d1n}, where the analyzed quantum number configurations are $J^P = 1/2^{\pm}$, $3/2^{\pm}$, and $5/2^-$ for $\bar{D}_1N$ and $1/2^-$, $3/2^{\pm}$, $5/2^{\pm}$, and $7/2^-$ for $\bar{D}_2^*N$. {To quantify the cutoff sensitivity, we present the binding energies as functions of the cutoff $\Lambda$ in Fig. \ref{TN}. We can see that for all promising molecular candidates, the binding energy varies around 10 MeV when $\Lambda$ changes by $\pm 0.1$GeV around the value where the state first appears.} 

For a cutoff $\Lambda \sim 1.00$ GeV, we can obtain bound-state solutions for the $\bar{D}_1N$ systems with $I(J^P) = 0(1/2^{\pm}, 3/2^+)$, $1(1/2^{\pm}, 3/2^+)$, and $0(3/2^-)$. The corresponding binding energies range from several to several tens of MeV, with rms radii on the order of 1.00 fm or larger, consistent with the properties of loosely bound molecular states. A comparison of cutoff values reveals a systematic pattern: for identical spin-parity with the $S$-wave interactions, the required cutoff is smaller for isoscalar states than for isovector states, e.g., $\Lambda(0(1/2^+)) < \Lambda(1(1/2^+))$ and $\Lambda(0(3/2^+)) < \Lambda(1(3/2^+))$. This indicates that the OBE effective potentials are more attractive in the isoscalar channels. Furthermore, $S$-wave states generally exhibit deeper binding than $P$-wave states with the same isospin and total spin, as evidenced by relations such as $\Lambda(0(1/2^+)) < \Lambda(0(1/2^-))$. As for the $\bar{D}_1N$ systems with $0(5/2^-)$ and $1(3/2^-, 5/2^-)$, no bound-state solutions are found within $\Lambda \leq 2.00$ GeV, indicating insufficient attraction in these channels.

Next, we perform a coupled-channel analysis for the $\bar{D}_1N/\bar{D}_2^*N$ systems. For the channels $I(J^P)=0(1/2^{\pm})$, $1(1/2^{\pm})$, $0(3/2^+)$, $1(3/2^+)$, and $0(3/2^-)$, the binding energies increase slightly compared to the single-channel $\bar{D}_1N$ results, while the qualitative features of the bound states remain unchanged. This confirms the molecular nature of these states and demonstrates that coupled-channel effects provide additional attractive contributions. For the systems with $1(5/2^-)$, bound states remain elusive even after considering the coupled-channel effects.

However, the coupled-channel calculation yields bound solutions for the $1(3/2^-)$ and $0(5/2^-)$ configurations. These states are dominated by the $\bar{D}_2^*N$ component and are characterized by rms radii smaller than 1.00 fm. Their compact size is inconsistent with a loosely bound molecular configuration. Therefore, they are not considered promising molecular candidates despite the presence of a sufficiently attractive potential.

Thus, we now focus on possible molecular states predominantly composed of $\bar{D}_2^*N$. As shown on the right half of Table \ref{d1n}, we find strong attractive interactions supporting loosely bound states for the $\bar{D}_2^*N$ systems with $1(3/2^-)$ and $0(5/2^-)$, with binding energies from several to tens of MeV, rms radii around or above 1.00 fm, and reasonable cutoff parameters, which all indicate that these two states qualify as promising molecular candidates.

In addition, we can identify six further $\bar{D}_2^*N$ molecular candidates with quantum numbers $I(J^P)=0(1/2^-, 3/2^{\pm}, 5/2^+)$ and $1(3/2^{+}, 5/2^+)$, all of which can yield loosely bound solutions within a reasonable cutoff range. Consistent with the $\bar{D}_1N$ findings, the analysis of cutoff values suggests that $S$-wave isoscalar configurations are the most favorable for molecular formation. For the $\bar{D}_2^*N$ systems with $0(7/2^-)$ and $1(1/2^-, 5/2^-, 7/2^-)$, no bound states can be found within $\Lambda \leq 2.00$ GeV, implying that the OBE potentials in these channels lack sufficient strength. 

In summary, we propose the following systems as promising anticharm molecular
pentaquark candidates:
\begin{itemize}
    \item $\bar{D}_1N$ states with $I(J^P) = 0,1(1/2^{\pm}, 3/2^+)$ and $0(3/2^-)$;
    \item $\bar{D}_2^*N$ states with $I(J^P) = 0,1(3/2^{\pm}, 5/2^+)$ and $0({1/2^-},5/2^-)$.
\end{itemize}

\subsection{The $\bar{D}_1\Lambda$, $\bar{D}_1\Sigma$, $\bar{D}_2^*\Lambda$ and $\bar{D}_2^*\Sigma$ systems}\label{secd1lam}

Compared to the $\bar{T}N$ systems, the $\bar{T}\Lambda$ interactions lack contributions from $\pi$ and $\rho$ meson exchange due to the isospin-forbidden $\Lambda\Lambda\pi(\rho)$ vertex. This absence of key short- and intermediate-range forces is expected to weaken the effective potential, making bound-state formation less likely in a single-channel analysis. For the $\bar{T}\Sigma$ systems, $\pi$ and $\rho$ exchange is permitted. However, the reduced number of light $u/d$ quarks in the $\Sigma$ hyperon relative to the nucleon may also lead to somewhat weaker OBE potentials compared to the $\bar{T}N$ case. Consequently, while single-channel $\bar{T}\Lambda$ molecules appear improbable, the introduction of coupled-channel effects with $\bar{T}\Sigma$ channels may yield viable $\bar{T}\Lambda/\bar{T}\Sigma$ molecular candidates dominated by the lower $\bar{T}\Lambda$ threshold, as well as single $\bar{T}\Sigma$ molecules.

\renewcommand\tabcolsep{0.32cm}
\renewcommand{\arraystretch}{1.7}
\begin{table*}[htpb]
\centering
\caption{The bound-state solutions  for the $\bar{T}\Lambda$ systems. Here, the units for the cutoff $\Lambda$, the binding energy $E$, and the root-mean-square radius $r_{rms}$ are GeV, MeV, and 
femtometer, respectively. Channels that are forbidden or dominant are marked with the symbols “$\cdots$” and in bold manner, respectively. The mass thresholds for the involved channels are  $m_{\bar{D}_1\Sigma}=3537.68$, $m_{\bar{D}_2^*\Sigma}=3578.73$, $m_{\bar{D}_1\Lambda}=3615.15$, and $m_{\bar{D}_2^*\Lambda}=3656.20$ MeV, respectively.}\label{D1L}
\begin{tabular}{ccccllll}
\toprule[1.5pt]
\toprule[1.5pt]

$ \quad I(J^P) \quad$  &$\Lambda$    &$E$    &$r_{rms}$      &$\bar{D}_1\Lambda(^{2}S/^{4}D)$   &$\bar{D}_2^*\Lambda(^{4}D/^{6}D)$
&$\bar{D}_1\Sigma(^{2}S/^{4}D)$   &$\bar{D}_2^*\Sigma(^{4}D/^{6}D)$\\ 
$\frac{1}{2}(\frac{1}{2}^+)$             &1.10   &$-$0.90        &3.67           &\textbf{84.19}     /0.16    &$\sim0$          /0.02    &15.06/0.55   &$\sim0$/0.02 \\
         &1.15   &$-$2.86        &2.01           &\textbf{62.51 }    /0.50    &$\sim0$          /0.05    &35.88/1.03   &$\sim0$/0.03  \\
\hline
   $I(J^P)$  &$\Lambda$    &$E$    &$r_{rms}$      &$\bar{D}_1\Lambda(^{4}S/^{2}D/^{4}D)$
 &$\bar{D}_2^*\Lambda(^{4}S/^{4}D/^{6}D)$ &$\bar{D}_1\Sigma(^{4}S/^{2}D/^{4}D)$
 &$\bar{D}_2^*\Sigma(^{4}S/^{4}D/^{6}D)$   \\ 
$\frac{1}{2}(\frac{3}{2}^+)$             &1.20   &$-$0.37        &5.13           &\textbf{96.66}  
/$\sim0$   /0.09&0.13/ $\sim0$/ $\sim0$
&1.28/1.77  /0.88  &0.78/$\sim0$  /0.01\\
                        &1.24  &$-$4.80  &1.68    &\textbf{65.21}/0.14/0.98           &1.04    /0.02    /0.15          &16.53  /0.61   / 3.28 &11.85   / 0.02  / 0.17       \\
                         \hline
$I(J^P)$  &$\Lambda$    &$E$    &$r_{rms}$      &$\bar{D}_1\Lambda(^{2}P/^{4}P)$
 &$\bar{D}_2^*\Lambda(^{4}P/^{6}P)$
 &$\bar{D}_1\Sigma(^{2}P/^{4}P)$
 &$\bar{D}_2^*\Sigma(^{4}P/^{6}P)$   \\  
$\frac{1}{2}(\frac{1}{2}^-)$
 &1.73        &$-$0.39           & 2.00        & 4.86/\textbf{67.29}&0.02/$\cdots$&17.07  / 10.29&0.47/$\cdots$   \\
 &1.75        &$-$6.05           & 1.11        & 4.87/\textbf{62.02}&0.02/$\cdots$&20.76  / 11.77&0.56/$\cdots$   \\ 
 $\frac{1}{2}(\frac{3}{2}^-)$
 &1.46        &$-$3.04           &1.16         &4.07/38.72&0.05/0.02  &6.48/\textbf{50.55}   &0.14/0.07     \\
 &1.47        &$-$8.10           &0.96         &3.63/35.50   &0.05/0.02  &6.78/\textbf{53.78}   &0.16/0.08     \\
 $\frac{1}{2}(\frac{5}{2}^-)$
 & 1.47       &$-$1.95           &0.73         &$\cdots$/$\sim0$   &6.86/25.70&$\cdots$/0.01     &17.25/\textbf{50.18}        \\
 & 1.48       &$-$9.25           &0.71         &$\cdots$/$\sim0$   &6.60/25.25&$\cdots$/$\sim0$     &17.23/\textbf{50.92}        \\
\midrule[1pt]\midrule[1pt]
   $I(J^P)$  &$\Lambda$    &$E$    &$r_{rms}$      &
 &$\bar{D}_2^*\Lambda(^{4}S/^{4}D/^{6}D)$ &$\bar{D}_1\Sigma(^{4}S/^{2}D/^{4}D)$
 &$\bar{D}_2^*\Sigma(^{4}S/^{4}D/^{6}D)$   \\ 
$\frac{1}{2}(\frac{3}{2}^+)$             &1.10   &$-$2.69       &2.22    &        &\textbf{74.28}  /0.05   /0.40&0.19/$\sim0$/ 0.01  &23.62/0.20/1.25
\\
                        &1.13  &$-$5.29  &1.55  & &\textbf{59.05}/0.10/0.71           &0.26    /$\sim0$    /0.01          &37.88  /0.27   / 1.72      \\ 
           \hline
   $I(J^P)$  &$\Lambda$    &$E$    &$r_{rms}$      &
 &$\bar{D}_2^*\Lambda(^{6}S/^{4}D/^{6}D)$ &
 &$\bar{D}_2^*\Sigma(^{6}S/^{4}D/^{6}D)$   \\ 
$\frac{1}{2}(\frac{5}{2}^+)$             &1.20   &$-$0.47       &4.89    &        &\textbf{97.44}  /0.03   /0.11&  &1.14/0.41/0.87
\\
                        &1.25  &$-$7.02  &1.53  & &\textbf{73.08}/0.43/1.28           &          &19.20  /1.81   / 4.20      \\ 
           \hline
$I(J^P)$  &$\Lambda$    &$E$    &$r_{rms}$      &
 &$\bar{D}_2^*\Lambda(^{4}P/^{6}P)$
 &$\bar{D}_1\Sigma(^{2}P/^{4}P)$
 &$\bar{D}_2^*\Sigma(^{4}P/^{6}P)$   \\  
$\frac{1}{2}(\frac{1}{2}^-)$
 &1.84        &$-$1.36           &1.25  &     &33.98/$\cdots$  & 4.39/0.88&\textbf{60.75}/$\cdots$   \\
 &1.86        &$-$5.87           &0.96     &    &29.79/$\cdots$  & 4.65/0.95&\textbf{64.61}/$\cdots$   \\ 
 $\frac{1}{2}(\frac{3}{2}^-)$
 &1.61   &$-$2.36       &0.87       &      &1.88  
/5.90   &10.33/\textbf{76.77}& 3.23/ 1.89
     \\
 &1.62       &$-$7.06           &0.82      &     &2.09/5.98   &10.15/\textbf{76.14}  &3.60/2.04      \\
  $\frac{1}{2}(\frac{5}{2}^-)$& 1.40       &$-$1.84           &1.33        &   &10.76/\textbf{35.17}   &0.02 
 &14.87  /\textbf{39.18}         \\
 & 1.41       &$-$6.53           &1.03        &   &9.61/\textbf{31.95}   &0.01 
 &15.93  /\textbf{42.50}         \\
   
$\frac{1}{2}(\frac{7}{2}^-)$             &$\cdots$   &$\cdots$       &$\cdots$    &        &$\cdots$/$\cdots$   &  &$\cdots$/$\cdots$
\\
\bottomrule[1.5pt]
\bottomrule[1.5pt]
\end{tabular}
\end{table*}

To verify the above analysis, a single-channel analysis of the $\bar{D}_1\Lambda$ and $\bar{D}_2^*\Lambda$ systems across the cutoff range $0.80 \leq \Lambda \leq 2.00$ GeV is performed 
yields no bound-stat solutions for any of the considered quantum numbers, confirming the insufficient attraction from $\sigma$ and $\omega$ exchange alone.

\begin{figure*}[!htpb]
\centering
\includegraphics[width=6.5in]{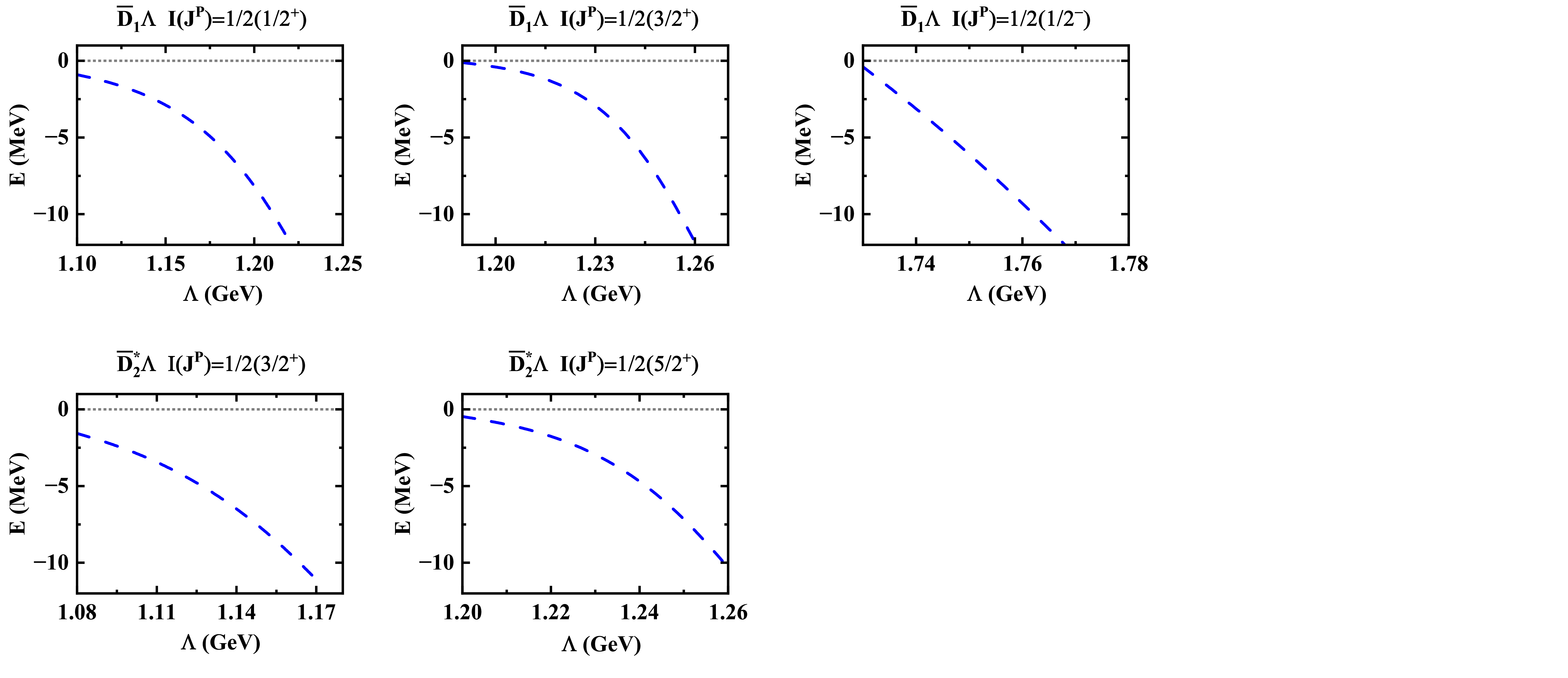}\\
\includegraphics[width=6.5in]{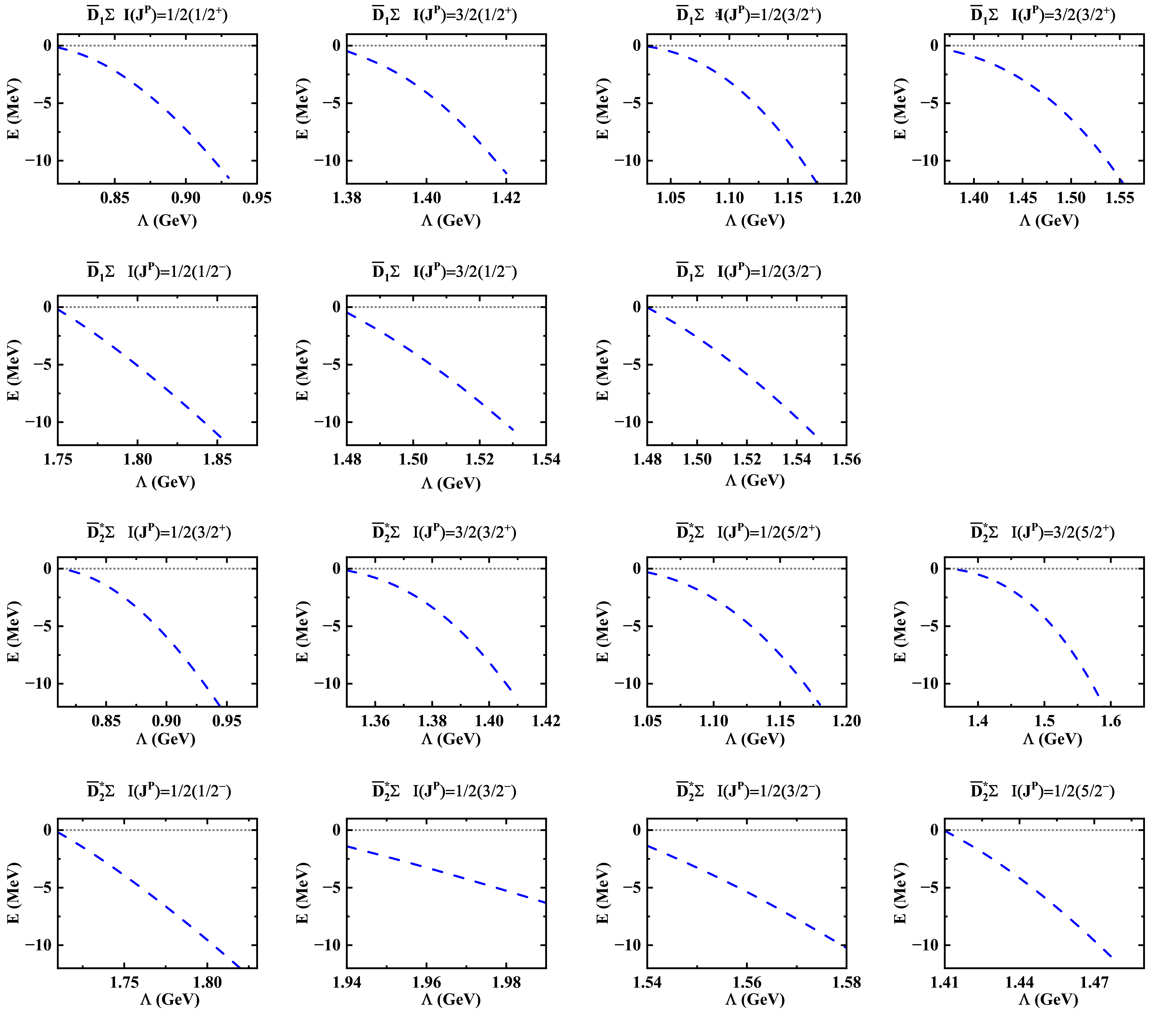}
\caption{The $\Lambda$ dependence of the binding energies $E$ for the $\bar{T}\Lambda$ and ${\bar{T}\Sigma}$ bound states.}
\label{TLTS}
\end{figure*}

However, bound states emerge when $\bar{T}\Sigma$ channels are coupled. As shown in Table \ref{D1L} and Fig. \ref{TLTS}, we find loosely bound solutions for the coupled $\bar{D}_1\Lambda/\bar{D}_2^*\Lambda/\bar{D}_1\Sigma/\bar{D}_2^*\Sigma$ systems with $I(J^P)=1/2(1/2^{\pm}, 3/2^{+})$ and the coupled $\bar{D}_2^*\Lambda/\bar{D}_1\Sigma/\bar{D}_2^*\Sigma$ systems with $1/2(3/2^{+}, 5/2^+)$.  For these states, the dominant component is the lowest $\bar{T}\Lambda$ channel. With binding energies of several MeV and rms radii on the order of 1.00 fm or larger, these systems satisfy the criteria for loosely bound molecules and are identified as promising molecular candidates. Their existence underscores the crucial role of coupled-channel effects in providing the necessary attraction.In addition, a quantitative analysis regarding the cutoff dependence of the bound-state solutions for the $\bar{D}_1\Lambda$ and  $\bar{D}_2^*\Lambda$ systems is presented in Figs. \ref{TLTS}. 

In contrast, for the $P$-wave coupled $\bar{D}_1\Lambda/\bar{D}_2^*\Lambda/\bar{D}_1\Sigma/\bar{D}_2^*\Sigma$ systems with $1/2(3/2^{-}, 5/2^-)$ and the $\bar{D}_2^*\Lambda/\bar{D}_1\Sigma/\bar{D}_2^*\Sigma$ systems with $1/2(1/2^{-}, 3/2^-, 5/2^-)$, bound solutions obtained near $\Lambda \sim 1.00$ GeV are dominated by the heavier $\bar{T}\Sigma$ components. This results in compact configurations with rms radii significantly below 1.00 fm, which are incompatible with a molecular interpretation. Finally, no bound-states are found for the $\bar{D}_2^*\Lambda/\bar{D}_1\Sigma/\bar{D}_2^*\Sigma$ system with $1/2(7/2^-)$ within the studied cutoff range.

\renewcommand\tabcolsep{0.35cm}
\renewcommand{\arraystretch}{1.8}
\begin{table*}[!htpb]
\centering
\caption{The bound-state solutions for the $\bar{T}\Sigma$ systems. Here, the units for the cutoff $\Lambda$, the binding energy $E$, and the root-mean-square radius $r_{rms}$ are GeV, MeV, and femtometer, respectively. Channels that are forbidden or dominant are marked with the symbols “$\cdots$” and in bold manner, respectively. The mass thresholds for the involved channels are    $m_{\bar{D}_1\Sigma}=3615.15$ and $m_{\bar{D}_2^*\Sigma}=3656.20$ MeV, respectively.}\label{D1S}
\begin{tabular}{ccccl|ccccl}
\toprule[1.5pt]
\toprule[1.5pt]
$ \quad I(J^P) \quad$  &$\Lambda$    &$E$    &$r_{rms}$   &$\bar{D}_1\Sigma(^{2}S/^{4}D)$  &$ \quad I(J^P) \quad$  &$\Lambda$    &$E$    &$r_{rms}$        &$\bar{D}_2^*\Sigma(^{4}S/^{4}D/^{6}D)$\\
 $\frac{1}{2}(\frac{1}{2}^+)$  &0.83   &$-$0.91&3.82&\textbf{99.60} /0.40              
  & $\frac{1}{2}(\frac{3}{2}^+)$ &0.85   &$-$1.43&3.18&\textbf{99.20} /0.11/0.68   
 \\
     &0.89     &$-$6.05    & 1.72  &\textbf{99.40}/0.60  & &0.90     &$-$5.91    & 1.75  &\textbf{98.91}/0.16/0.93   
     \\ 
      $\frac{3}{2}(\frac{1}{2}^+)$  &1.38   &$-$0.46&4.70&\textbf{97.68} /2.32                
     &$\frac{3}{2}(\frac{3}{2}^+)$  &1.35   &$-$0.13&5.91&\textbf{98.16}/0.35/1.49      \\
     &1.41     &$-$7.02    & 1.45  &\textbf{93.92}/6.08  & &1.39     &$-$5.34    &  1.71  &\textbf{93.25} /1.34/5.41    
    \\ \hline
   $I(J^P)$  &$\Lambda$    &$E$    &$r_{rms}$      &$\bar{D}_1\Sigma(^{4}S/^{2}D/^{4}D)$  &$ \quad I(J^P) \quad$   &$\Lambda$    &$E$    &$r_{rms}$      
 &$\bar{D}_2^*\Sigma(^{6}S/^{4}D/^{6}D)$   \\ 
$\frac{1}{2}(\frac{3}{2}^+)$    &1.05   &$-$0.49       &4.91           &\textbf{98.50}  
/0.25   /1.25                     
& $\frac{1}{2}(\frac{5}{2}^+)$   & 1.10  &$-$2.56  &2.65  &\textbf{96.88}/0.99/2.13
      \\
            & 1.15  &$-$8.29  &1.63  &\textbf{96.10}/0.66/3.24
            &  & 1.13  &$-$5.15  &1.98  &\textbf{96.06}/1.24/2.70            
                        \\
            $\frac{3}{2}(\frac{3}{2}^+)$    &1.40   &$-$0.96       &3.83           &\textbf{96.24}  
/0.89   /2.87                     
& $\frac{3}{2}(\frac{5}{2}^+)$   &1.40  &$-$0.49  &4.78  &\textbf{97.47}/1.05/1.48
   \\
            & 1.50 &$-$6.36  &1.76  &\textbf{90.91}/2.22/6.87
             &  &1.50  &$-$4.22  &2.08  &\textbf{93.22}/2.89/3.89          
                        \\\toprule[1.5pt]
$I(J^P)$  &$\Lambda$    &$E$    &$r_{rms}$      &$\bar{D}_1\Sigma(^{2}P/^{4}P)$   &$ \quad I(J^P) \quad$  &$\Lambda$    &$E$    &$r_{rms}$    
 &$\bar{D}_2^*\Sigma(^{4}P/^{6}P)$   \\ 
 $\frac{1}{2}(\frac{1}{2}^-)$    &1.75    &$-$0.19    &2.65    &\textbf{82.55}/17.45      
&$\frac{1}{2}(\frac{1}{2}^-)$   &1.72  &$-$1.02  &1.99  &\textbf{100.00}   /$\cdots$ 
 \\
 & 1.80     &$-$5.08    &1.36    &\textbf{82.30}/17.00   &&1.77  &$-$6.06  &1.30  &\textbf{100.00}   /$\cdots$  
 \\ 
 $\frac{3}{2}(\frac{1}{2}^-)$& 1.48     &$-$0.45    &2.60    &17.76/\textbf{82.24}      
 &$\frac{1}{2}(\frac{3}{2}^-)$     &1.94      &$-$1.41    &1.80     &\textbf{68.64 }/31.36 
 \\
&1.52     &$-$8.20    &1.31    &18.16/\textbf{81.84}      & &1.98      &$-$5.26    &1.31     &\textbf{68.90} /31.10  
     \\
  $\frac{1}{2}(\frac{3}{2}^-)$     &1.48      &$-$0.02    &3.31     &11.75 /\textbf{88.25}       
 & $\frac{3}{2}(\frac{3}{2}^-)$   &1.54  &$-$1.35  &2.00  &32.37   /\textbf{67.63} 
 \\
 &1.53      &$-$7.63    &1.33     &12.31 /\textbf{87.69}   & &1.57  &$-$7.67  &1.30  &33.08   /\textbf{66.92 }  
      \\  
  $\frac{3}{2}(\frac{3}{2}^-)$     &$\cdots$      &$\cdots$    &$\cdots$     &$\cdots$ /$\cdots$      
 & $\frac{1}{2}(\frac{5}{2}^-)$   &1.41  &$-$0.05  &3.31  &26.69   /\textbf{73.31} 
 \\
     
$\frac{1}{2}(\frac{5}{2}^-)$     &$\cdots$      &$\cdots$    &$\cdots$     &$\cdots$ /$\cdots$       
 &     &1.45  &$-$5.81  &1.46  &28.02   /\textbf{71.98 }  
 \\
   
 $\frac{3}{2}(\frac{5}{2}^-)$     &$\cdots$      &$\cdots$    &$\cdots$     &$\cdots$ /$\cdots$      
  & $\frac{3}{2}(\frac{1}{2}^-)$     &$\cdots$      &$\cdots$    &$\cdots$     &$\cdots$ /$\cdots$ 
 \\
&      &    &     & & $\frac{3}{2}(\frac{5}{2}^-)$     &$\cdots$      &$\cdots$    &$\cdots$     &$\cdots$ /$\cdots$ 
 \\
  &      &    &     &       
  &$\frac{1}{2}(\frac{7}{2}^-)$    &$\cdots$  &$\cdots$  &$\cdots$  &$\cdots$   /$\cdots$ \\
  &      &    &     &       
  &$\frac{3}{2}(\frac{7}{2}^-)$    &$\cdots$  &$\cdots$  &$\cdots$  &$\cdots$   /$\cdots$ \\
\bottomrule[1.5pt]
\bottomrule[1.5pt]
\end{tabular}
\end{table*}

The bound-state solutions for the $\bar{D}_1N$ and $\bar{D}_2^*N$ systems, incorporating $S$-$D$ wave mixing and $P$-wave interactions, are summarized in Table \ref{d1n}, where the analyzed quantum number configurations are $J^P = 1/2^{\pm}$, $3/2^{\pm}$, and $5/2^-$ for $\bar{D}_1N$, and $1/2^-$, $3/2^{\pm}$, $5/2^{\pm}$, and $7/2^-$ for $\bar{D}_2^*N$.

We now investigate possible molecular states dominated by $\bar{T}\Sigma$ components. The bound-state solutions for single-channel $\bar{D}_1\Sigma$ and $\bar{D}_2^*\Sigma$ systems are collected in Table \ref{D1S} for $0.80 \leq \Lambda \leq 2.00$ GeV, where loosely bound states are found as
\begin{itemize}
    \item $\bar{D}_1\Sigma$ states with $I(J^P) = 1/2, 3/2(1/2^{\pm}, 3/2^+)$ and $1/2(3/2^-)$;
    \item $\bar{D}_2^*\Sigma$ states with $I(J^P) = 1/2, 3/2(3/2^+, 5/2^+)$, $1/2(1/2^-, 5/2^-)$, and $1/2, 3/2(3/2^-)$.
\end{itemize}
These states, characterized by reasonable cutoff values, binding energies of several MeV, and rms radii around 1.00 fm, qualify as good molecular candidates.

A subsequent coupled-channel analysis of the $\bar{D}_1\Sigma/\bar{D}_2^*\Sigma$ systems shows that the binding energies increase modestly, but the qualitative features of the states remain largely unchanged. This indicates that while coupled-channel effects provide a positive, attractive contribution, they are not essential for the formation of these particular $\bar{T}\Sigma$ molecules. 

For $P$-wave $\bar{D}_1\Sigma$ with $1/2(5/2^-)$, $3/2(3/2^-, 5/2^-)$ or $\bar{D}_2^*\Sigma$ with $1/2(7/2^-)$, $3/2(1/2^-, 5/2^-, 7/2^-)$, no single-channel bound states can be found. However, when the coupled-channel effects are further introduced, bound solutions appear for the $\bar{D}_1\Sigma/\bar{D}_2^*\Sigma$ systems with $1/2(5/2^-)$ and $3/2(3/2^-)$. Unfortunately, in these cases, the $\bar{D}_1\Sigma$ component is not dominant, and the resulting rms radii deviate significantly from 1.00 fm. Therefore, these coupled states are not considered suitable molecular candidates.

\subsection{The $\bar{D}_1\Xi$ and $\bar{D}_2^*\Xi$ systems}

\begin{table*}[!htpb]
\renewcommand\tabcolsep{0.32cm}
\renewcommand{\arraystretch}{1.7}
\centering
\caption{The bound-state solutions for the $\bar{T}\Xi$ systems after considering the $S-D$-wave mixing effects and the $P$-wave interactions. Here, the units for the cutoff $\Lambda$, the binding energy $E$, and the root-mean-square radius $r_{rms}$ are GeV, MeV, and femtometer, respectively. Channels that are forbidden or dominant are marked with the symbols “$\cdots$” and in bold manner, respectively. The mass thresholds for the involved channels are $m_{\bar{D}_1\Xi}=3740.29$ , and $m_{\bar{D}_2^*\Xi}=3781.34$ MeV, respectively.}\label{D1X}
\begin{tabular}{ccccc|ccccccccc}
\toprule[1.5pt]
\toprule[1.5pt]
$ \quad I(J^P) \quad$  &$\Lambda$    &$E$    &$r_{RMS}$   &$\bar{D}_1\Xi(^{2}S/^{4}D)$  &$ \quad I(J^P) \quad$ &$\Lambda$    &$E$    &$r_{RMS}$       &$\bar{D}_2^*\Xi(^{4}S/^{4}D/^{6}D)$\\
 $0(\frac{1}{2}^+)$  &1.10   &$-$2.96&2.33&\textbf{99.80} /0.20                & $0(\frac{3}{2}^+)$  &1.06   &$-$0.76&4.13&\textbf{99.81} /0.03/0.16   \\
     &1.11     &$-$3.96    &  2.06  &\textbf{99.78}/0.22  & &1.12     &$-$5.91    & 1.74  &\textbf{99.63}/0.06/0.31    
     \\ 
  $1(\frac{1}{2}^+)$  &$\cdots$      &$\cdots$    &$\cdots$     &$\cdots$ /$\cdots$                & $1(\frac{3}{2}^+)$      &$\cdots$    &$\cdots$     &$\cdots$ /$\cdots$  &$\cdots$ /$\cdots$  \\\hline
   $I(J^P)$  &$\Lambda$    &$E$    &$r_{RMS}$      &$\bar{D}_1\Xi(^{4}S/^{2}D/^{4}D)$  &$I(J^P)$  &$\Lambda$    &$E$    &$r_{RMS}$      
 &$\bar{D}_2^*\Xi(^{6}S/^{4}D/^{6}D)$   \\ 
$0(\frac{3}{2}^+)$   & 1.00  &$-$1.02  &3.66  &\textbf{99.74}/0.05/0.21                     &$0(\frac{5}{2}^+)$   & 1.01   &$-$1.87&2.83&\textbf{99.69}/0.11/0.20    \\
            & 1.05  &$-$4.16  &2.03  &\textbf{99.60}/0.08/0.32
                        &  &1.09     &$-$8.64    &  1.51  &\textbf{99.52} /0.18/0.30
                        \\ 
            $1(\frac{3}{2}^+)$   &1.70  &$-$0.81  &3.95  &\textbf{98.88}/0.18/0.94      
                  &$1(\frac{5}{2}^+)$   &1.70  &$-$1.58  &2.98  &\textbf{98.32}/0.51/1.17        \\
           & 1.90  &$-$8.47  &1.44  &\textbf{96.28}/0.55/3.17
                       &&1.80  &$-$5.50  &1.72  &\textbf{96.78}/0.95/2.27  
                       \\\toprule[1.5pt]
$I(J^P)$  &$\Lambda$    &$E$    &$r_{RMS}$      &$\bar{D}_1\Xi(^{2}P/^{4}P)$  &$I(J^P)$  &$\Lambda$    &$E$    &$r_{RMS}$     
 &$\bar{D}_2^*\Xi(^{4}P/^{6}P)$   \\
 $0(\frac{1}{2}^-)$    &1.67    &$-$0.65    &2.25    &18.23/\textbf{81.77}      
 & $0(\frac{3}{2}^-)$   &1.70  &$-$0.92  &2.05  &33.39   /\textbf{66.61}
 \\
 &1.72    &$-$7.17    &1.25    &19.29/\textbf{80.71}      
 &  &1.75  &$-$7.82  &1.21  &35.19   /\textbf{64.81} 
 \\ 
 $1(\frac{1}{2}^-)$& $\cdots$     &$\cdots$    &$\cdots$    &$\cdots$/$\cdots$      
 & $0(\frac{1}{2}^-)$& $\cdots$    &$\cdots$    &$\cdots$  &$\cdots$/$\cdots$          \\
  $0(\frac{3}{2}^-)$     &$\cdots$     &$\cdots$    &$\cdots$    &$\cdots$/$\cdots$ 
 & $1(\frac{1}{2}^-)$ &$\cdots$     &$\cdots$    &$\cdots$    &$\cdots$/$\cdots$  
 
 \\
  $1(\frac{3}{2}^-)$     &$\cdots$      &$\cdots$    &$\cdots$     &$\cdots$ /$\cdots$      
 &$1(\frac{3}{2}^-)$      &$\cdots$    &$\cdots$    &$\cdots$ &$\cdots$/$\cdots$ \\
$0(\frac{5}{2}^-)$     &$\cdots$      &$\cdots$    &$\cdots$     &$\cdots$/$\cdots$   
  &$0(\frac{5}{2}^-)$      &$\cdots$    &$\cdots$     &$\cdots$   &$\cdots$/$\cdots$     \\
 $1(\frac{5}{2}^-)$     &$\cdots$      &$\cdots$    &$\cdots$     &$\cdots$ /$\cdots$      &$1(\frac{5}{2}^-)$        &$\cdots$           &$\cdots$         &$\cdots$   &$\cdots$/$\cdots$     \\ &      &    &     &       
  &$0(\frac{7}{2}^-)$    &$\cdots$  &$\cdots$  &$\cdots$  &$\cdots$   /$\cdots$ \\
 &      &    &     &       
  &$1(\frac{7}{2}^-)$    &$\cdots$  &$\cdots$  &$\cdots$  &$\cdots$   /$\cdots$ \\ 
\bottomrule[1.5pt]
\bottomrule[1.5pt]
\end{tabular}
\end{table*}

\begin{figure*}[!htpb]
\centering
\includegraphics[width=6.5in]{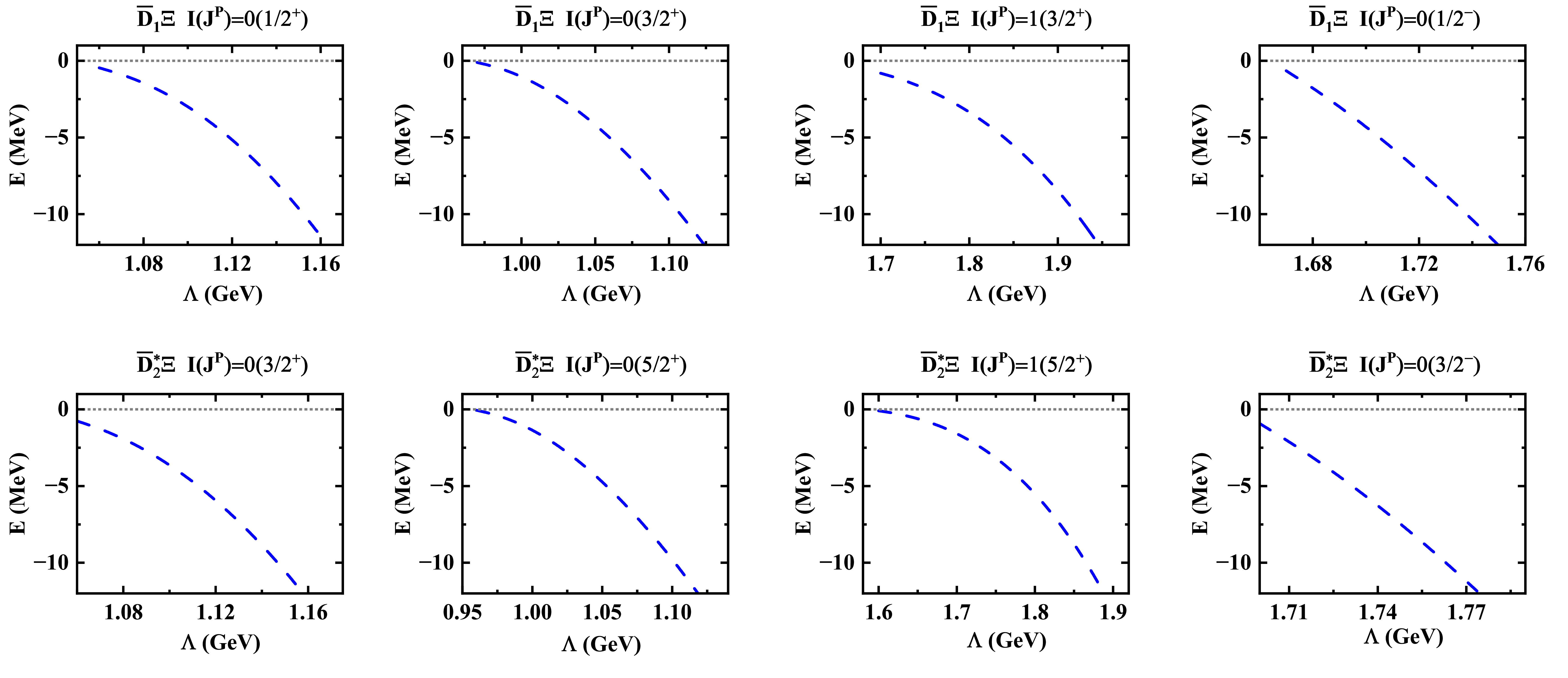}
\caption{The $\Lambda$ dependence of the binding energies $E$ for the ${{\bar{T}\Xi}}$ bound states.}
\label{TX}
\end{figure*}

For the $\bar{T}\Xi$ systems, the reduced mass is somewhat larger than for $\bar{T}N$, which makes it a factor that generally favors binding. However, the OBE effective potentials are expected to be weaker due to the reduced number of light $u/d$ quarks in the $\Xi$ hyperon, diminishing the effects from light meson exchange.

The bound-state solutions for the $\bar{T}\Xi$ systems, incorporating $S$-$D$-wave mixing and $P$-wave interactions, are presented in Table \ref{D1X} and Fig. \ref{TX}.  For a cutoff $\Lambda \sim 1.00$ GeV, we obtain loosely bound solutions for the $\bar{D}_1\Xi$ states with $I(J^P)=0(1/2^{\pm}, 3/2^+)$ and $1(3/2^+)$ and the $\bar{D}_2^*\Xi$ states with $0(3/2^{\pm}, 5/2^+)$ and $1(5/2^+)$. A comparison with the $\bar{T}N$ systems reveals that, for identical binding energies, the required cutoff values are slightly larger for the corresponding $\bar{T}\Xi$ states. This cutoff hierarchy confirms that the OBE potentials are indeed somewhat less attractive in the $\bar{T}\Xi$ sector. The rms radii of the states listed above are consistent with the typical size of a loosely bound molecule. Therefore, we identify them as promising molecular candidates.

Furthermore, we observe that, for the positive-parity bound states, the $S$-wave channel provides the dominant component. In contrast, for the negative-parity states, the higher-spin channels exhibit the largest probabilities.

While for the $\bar{D}_1\Xi$ systems with $0(3/2^-, 5/2^-)$ and $1(1/2^+, 1/2^-, 3/2^-, 5/2^-)$, and the $\bar{D}_2^*\Xi$ systems with $0(1/2^-, 5/2^-, 7/2^-)$ and $1(1/2^-, 3/2^+, 3/2^-, 5/2^-, 7/2^-)$, no bound-state solutions are found within a reasonable cutoff range. This indicates that the OBE effective potentials for these channels lack sufficient strength to form molecular states. This conclusion remains unchanged even after including coupled-channel effects for the $\bar{D}_1\Xi/\bar{D}_2^*\Xi$ systems with $0(3/2^-, 5/2^-)$ and $1(1/2^+, 1/2^-, 3/2^-, 5/2^-)$.

Here, we also notice there are additional channel families, such as the $D_{s1}N$, $D_{s2}^*N$, $D_{s1}\Lambda(\Sigma)$, and $D_{s2}^*\Lambda(\Sigma)$, with their mass threshold close to the coupled $\bar{T}B$ systems with $|S| = 1, 2$. When we further consider the contributions from these additional channels, we cannot obtain additional loosely bound states composed by a charmed meson and a light baryon, but find that there can exist possible bound states dominantly composed by a charm-strange meson and a light baryon. For example, as shown in Table \ref{D1X}, there cannot appear bound-state solutions for the couple $\bar{D}_1\Xi/\bar{D}_2^*\Xi$ systems with $I(J^P)=1(3/2^{-}, 5/2^-)$. If we perform a coupled-channel analysis on the $\bar{D}_1\Xi/\bar{D}_{s2}^*\Sigma/\bar{D}_2^*\Xi$ systems with $1(3/2^-, 5/2^-)$, although we can obtain bound state solutions with $\Lambda\leq2.00$ GeV, the channels with higher mass play a important role, which makes the rms radii for these bound states a little smaller for a loosely bound molecular candidates, therefore, they cannot be good molecular candidates. In a future study, we will further explore the existence of possible molecular candidates composed by a charm-strange meson and a light baryon.

\section{Conclusion and discussion}\label{sec4}

\begin{figure*}[!h]
\centering
$\left.\begin{array}{lll}
\includegraphics[width=6.8in]{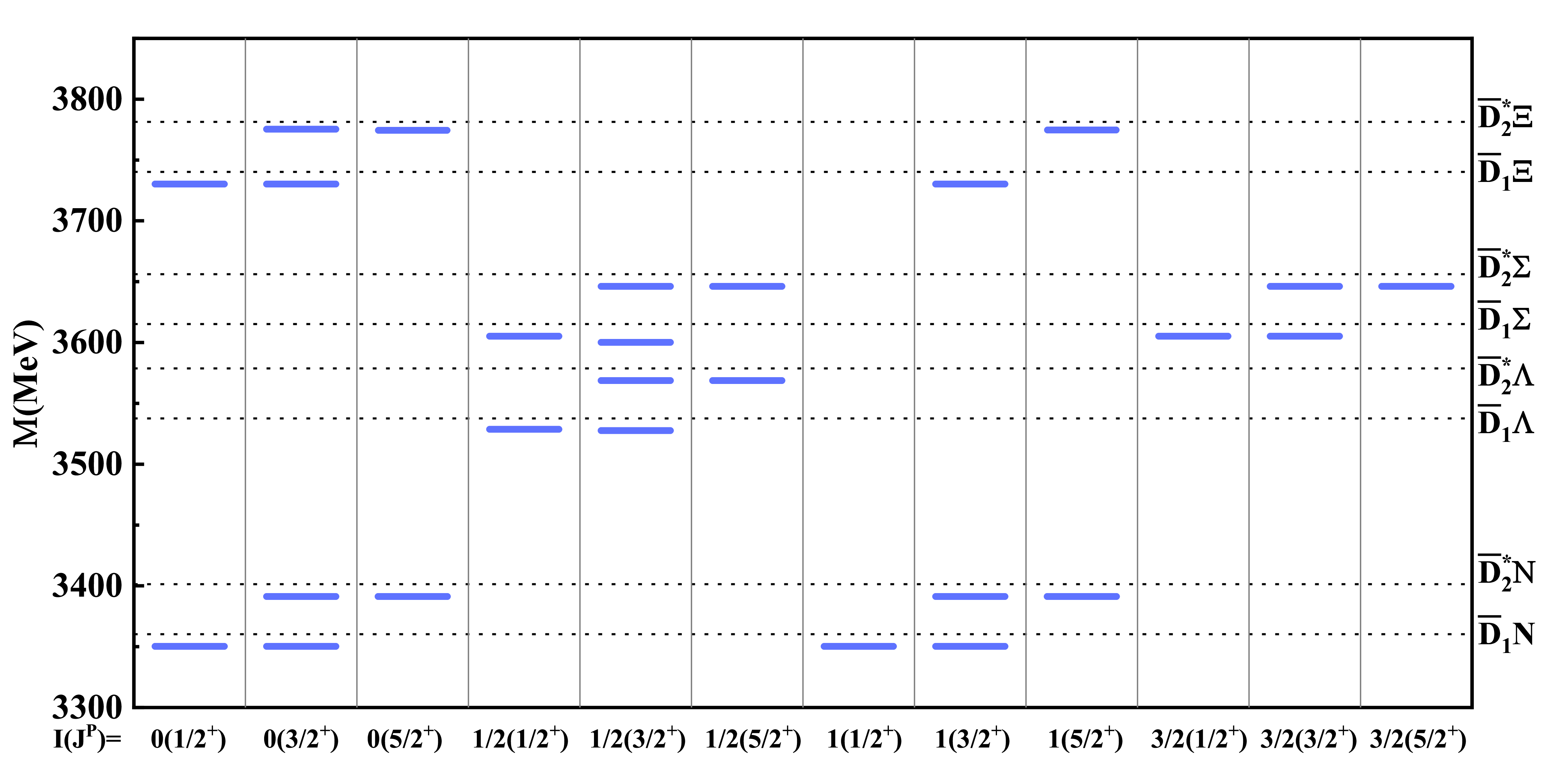}\\
\includegraphics[width=6.8in]{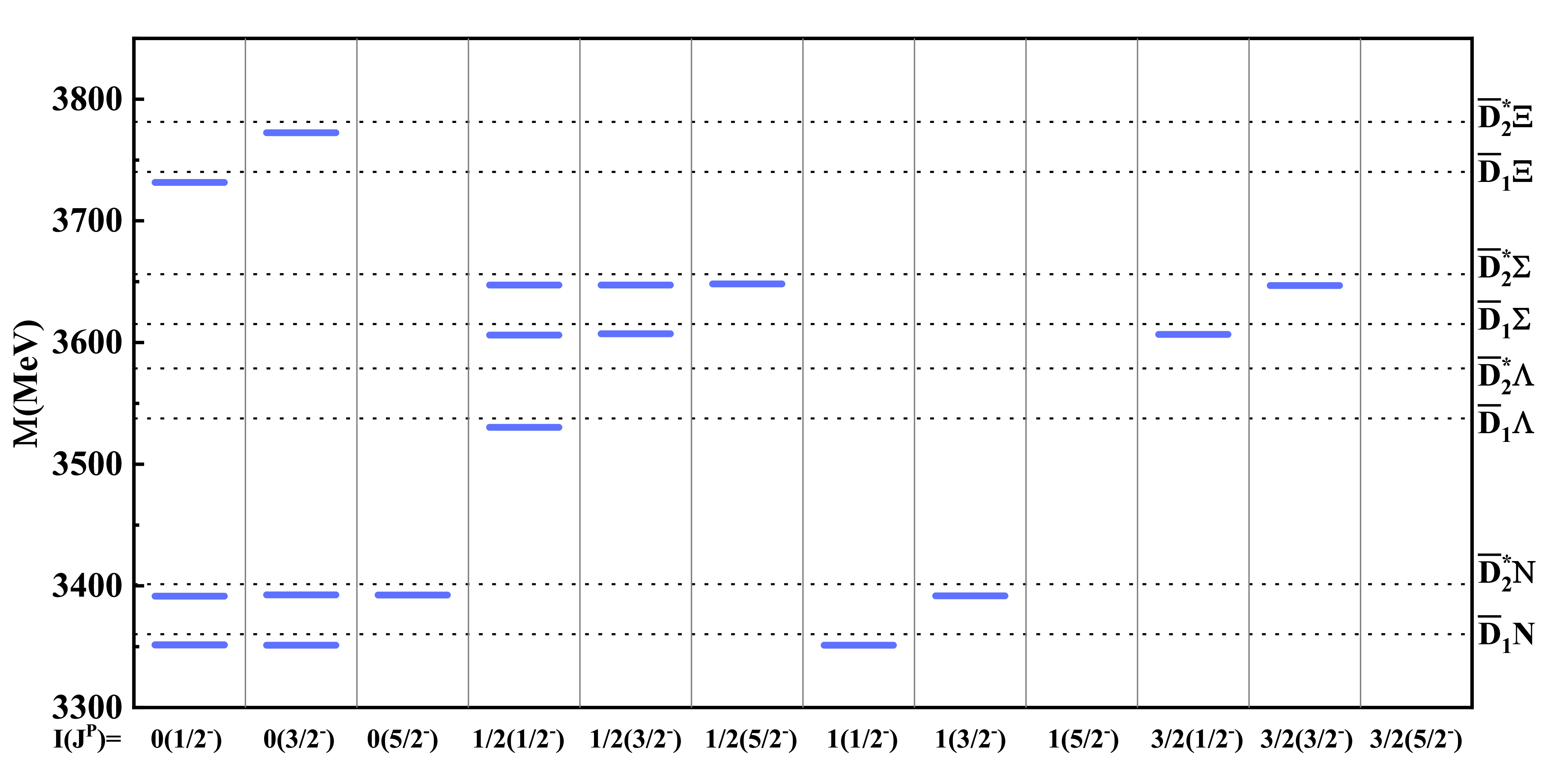}
\end{array}\right.$
\caption{A summary of predictions of possible molecular pentaquarks. We show here, the (upper)  positive-parity pentaquark molecular states and the (lower) negative-parity ones. The short solid  and the long dashed lines label the mass locations for possible molecular candidates and the mass thresholds for  the anticharm meson and light baryon systems, respectively.}
\label{summary1}
\end{figure*}

The study of pentaquarks is a rapidly evolving field that has seen significant progress in recent years, particularly in the identification of new exotic hadrons at high-energy colliders. While many pentaquark states have been observed experimentally, their detailed structure and theoretical models are still being developed. This area of research continues to be one of the most exciting frontiers in hadron physics.

In this work, we have systematically investigated the interactions between the $P$-wave anticharmed mesons $(\bar{D}_1, \bar{D}_2^*)$ and the light octet baryons $(N, \Lambda, \Sigma, \Xi)$ within the framework of the OBE model. Our analysis incorporates the coupled-channel effects, the $S$-$D$-wave mixing, and $P$-wave interactions. The relevant coupling constants are determined by extending the well-established nucleon-nucleon interaction parameters via the quark model and $SU(3)$ symmetry.

By numerically solving the coupled-channel Schr\"{o}dinger equations, we can predict a series of promising anticharmed molecular pentaquark candidates with strangeness $|S| = 0, 1, 2$. As summarized in Fig. \ref{summary1}, these candidates include
\begin{enumerate}
    \item $\bar{T}N$ molecules: $\bar{D}_1N$ states with $I(J^P)=0(1/2^{\pm}, 3/2^{\pm})$, $1(1/2^{\pm}, 3/2^+)$; $\bar{D}_2^*N$ states with $0(1/2^-, 3/2^{\pm}, 5/2^{\pm})$, $1(3/2^{\pm}, 5/2^+)$.
    \item $\bar{T}\Sigma$ molecules: $\bar{D}_1\Sigma$ states with $1/2(1/2^{\pm}, 3/2^{\pm})$, $3/2(1/2^{\pm}, 3/2^{+})$; $\bar{D}_2^*\Sigma$ states with $1/2(1/2^-, {3/2^{\pm}}, 5/2^{\pm})$, $3/2(3/2^{\pm}, 5/2^+)$; $\bar{D}_1\Lambda/\bar{D}_2^*\Lambda/\bar{D}_1\Sigma/\bar{D}_2^*\Sigma$ coupled states with $1/2(1/2^{\pm}, 3/2^{+})$ and $\bar{D}_2^*\Lambda/\bar{D}_1\Sigma/\bar{D}_2^*\Sigma$ coupled states with $1/2(3/2^{+}, 5/2^+)$.
    \item $\bar{T}\Xi$ molecules: $\bar{D}_1\Xi$ states with $0(1/2^{\pm}, 3/2^+)$, $1(3/2^+)$; $\bar{D}_2^*\Xi$ states with $0(3/2^{\pm}, 5/2^+)$, $1(5/2^+)$.
\end{enumerate}

{Furthermore, both $D_1$ and $D_2^*$ mesons have a width of tens of MeV. These non-negligible widths may affect the near-threshold poles and the  corresponding experimental signatures. In fact, the instability of constituent particles can introduce three-body effects, primarily comprising the self-energy and the OPE recoil corrections. While these effects are well studied in narrow systems like $T_{cc}$ and $X(3872)$ \cite{Du:2021zzh}, a recent study \cite{Cheng:2026cgo} demonstrated that for broad constituents like $\bar{D}_1$ and $\bar{D}_2^*$, the OPE recoil correction is strongly suppressed, leaving the self-energy correction dominant. Taking the $B D_1$ system as an example, this dominant self-energy effect transforms the near-threshold bound state into a quasibound state. This state locates on the first Riemann sheet with respect to the two-body $B D_1$ threshold, but on the second Riemann sheet of the three-body $B D^*\pi$ threshold. Consequently, the real energy of the pole experiences only a minor shift, while it acquires an imaginary part comparable to the half-width of the $D_1$ meson. This effectively shifts the two-body threshold into the complex plane: $m_{th} \to m_1 + m_2 - i\Gamma_{D_1}/2$.}

Regarding experimental observability, although this quasibound state ultimately decays into the three-body $B D^*\pi$ channel, its invariant mass spectrum will not exhibit a symmetric Breit-Wigner peak due to its proximity to the $B D_1$ threshold. Instead, it manifests as a distinctly asymmetric threshold enhancement: the lower-energy side displays a broad tail governed by the large decay width of the $D_1$ meson, while the higher-energy side is sharply truncated by the opening of the two-body $B D_1$ phase space.

In addition, our analysis demonstrates that the coupled-channel effects contribute actively to the formation of single-channel molecular states and play an important role in the formation of coupled-channel molecular states. Thus, experimental search for the molecular pentaquarks predicted in this work can not only test the molecular paradigm, but also can deepen our understanding on the nonperturbative interaction and coupled-channel effects in multiquark systems.

\section*{ACKNOWLEDGMENTS}

This project is supported by the National Natural Science Foundation of China under Grants No. 12305139 and No. 12305087 and the Xiaoxiang Scholars Programme of Hunan Normal University.

\end{document}